\documentclass[12pt,a4paper]{article}
\pdfoutput=1
\usepackage{jheppub}
\usepackage{graphicx}
\usepackage{lipsum}
\usepackage[all]{hypcap}
\allowdisplaybreaks

\newcommand{\KL}{K\"{a}ll\'{e}n-Lehmann}

\usepackage{xspace}
\newcommand{\MadGraph}{{\rmfamily\scshape MadGraph5\_aMC@NLO}\xspace}
\newcommand{\FeynRules}{{\rmfamily\scshape FeynRules}\xspace}
\newcommand{\UFO}{{\rmfamily\scshape UFO}\xspace}

\newcommand{\be}{\begin{equation}}
\newcommand{\ee}{\end{equation}}
\newcommand{\bea}{\begin{eqnarray}}
\newcommand{\eea}{\end{eqnarray}}
\newcommand{\beq}{\begin{equation}}
\newcommand{\eeq}{\end{equation}}

\def\Eq#1{eq.~(\ref{#1})}
\def\Fig#1{fig.~\ref{#1}}
\def\Sec#1{sect.~\ref{#1}}
\def\App#1{appendix~\ref{#1}}
\def\Tab#1{table~\ref{#1}}

\begin{document}
%%%%%%%%%%%%%%%%%%%%%%%%%%%%%%%%%%%%%%%%%%%%%%%%%%%%%%%%%%%%%%%%%%%%%%
%\preprint{ULB-TH/17-??}
{\raggedleft CERN-TH-2019-027 \\}
{\raggedleft DAMTP-2019-8 \\}

\title{The {\boldmath $\hat{H}$-Parameter}: An Oblique Higgs View}

\author[a]{Christoph Englert,}
\emailAdd{christoph.englert@cern.ch}

\author[b]{Gian F. Giudice,}
\emailAdd{gian.giudice@cern.ch}

\author[b]{Admir Greljo,}
\emailAdd{admir.greljo@cern.ch}

\author[b,c]{Matthew McCullough}
\emailAdd{matthew.mccullough@cern.ch}

\affiliation[a]{SUPA, School of Physics \& Astronomy, University of Glasgow, Glasgow, UK}

\affiliation[b]{CERN, Theoretical Physics Department, Geneva, Switzerland}

\affiliation[c]{DAMTP, University of Cambridge, Wilberforce Road, Cambridge, UK}

%%%%%%%%%%%%%%%%%%%%%%%%%%%%%%%%%%%%%%%%%%%%%%%%%%%%%%%%%%%%%%%%%%%%%%
\abstract{We study, from theoretical and phenomenological angles, the Higgs boson oblique parameter $\hat{H}$, as the hallmark of off-shell Higgs physics.  $\hat{H}$ is defined as the Wilson coefficient of the sole dimension-6 operator that modifies the Higgs boson propagator, within a Universal EFT.  Theoretically, we describe self-consistency conditions on Wilson coefficients, derived from the \KL~representation.    Phenomenologically, we demonstrate that the process $gg\to h^\ast \to VV$ is insensitive to propagator corrections from $\hat{H}$, and instead advertise four-top production as an effective high-energy probe of off-shell Higgs behaviour, crucial to break flat directions in the EFT.}

\maketitle
%%%%%%%%%%%%%%%%%%%%%%%%%%%%%%%%%%%%%%%%%%%%%%%%%%%%%%%%%%%%%%%%%%%%%%

\section{Introduction}
Oblique corrections to gauge boson propagators have played a prominent role in the analysis of electroweak precision data~\cite{Peskin:1990zt,Golden:1990ig,Holdom:1990tc,Altarelli:1990zd,Grinstein:1991cd,Peskin:1991sw,Altarelli:1991fk}. In an effective field theory (EFT) context, at invariant momenta $q^2$ smaller than the heavy new-physics mass scale (here denoted by $M$) the self-energy of electroweak (EW) gauge bosons can be expanded as
\be
\Pi_V (q^2) =  \Pi_V (0) + q^2 \Pi^\prime_V (0) + \frac{q^4}{2} \Pi^{\prime \prime}_V (0) + \dots
\ee
where the primes denote derivatives with respect to $q^2$.  When the expansions are truncated at order $q^4$~\cite{Burgess:1993mg,Maksymyk:1993zm,Barbieri:2004qk}, the leading electroweak oblique corrections are fully described by only 4 parameters, called $\hat{S}$, $\hat{T}$, $\hat{W}$, $\hat{Y}$.\footnote{Usually $\hat{W}$ and $\hat{Y}$ are called simply $W$ and $Y$, but we prefer a notation that avoids confusion between oblique parameters and gauge fields or hypercharge.} These parameters contribute to physical amplitudes at different orders in $q^2$. In particular, one finds $\hat{T} = {\mathcal O}(q^0)$, $\hat{S} = {\mathcal O}(q^2)$, and $\hat{W},\hat{Y} = {\mathcal O}(q^4)$.  This explains why $\hat{S}$ and $\hat{T}$ are the key parameters for LEP1 analyses, while $\hat{W}$ and $\hat{Y}$ play a critical role when LEP2 data are considered~\cite{Barbieri:2004qk}.  Recently the $\hat{W}$ and $\hat{Y}$ parameters have received renewed attention, due to the fact that their energy-growing contribution to amplitudes can be strongly constrained at high energy hadron colliders, allowing for precision EW probes at the LHC and beyond~\cite{Farina:2016rws,Franceschini:2017xkh,Banerjee:2018bio}.

In this work we focus on ${\mathcal O}(q^4)$ terms and, since the Higgs boson has now become a core component of the electroweak sector, we seek to add the Higgs analogue of the $\hat{W}$ and $\hat{Y}$ parameters, the $\hat{H}$-parameter, to the oblique dictionary.\footnote{Here we are focusing on the self-energy of the real Higgs boson, while the other three components of the Higgs doublet, which form the longitudinal gauge degrees of freedom, were already partly included in the EW oblique parameters.}  Defined within a dimension-6 EFT, the $\hat{W}$, $\hat{Y}$, and $\hat{H}$ parameters are
\be
\mathcal{L}_{\hat{W}} = -\frac{\hat{W}}{4 m_W^2} (D_\rho W^a_{\mu\nu})^2 ~~,~~\mathcal{L}_{\hat{Y}} = -\frac{\hat{Y}}{4 m_W^2} (\partial_\rho B_{\mu\nu})^2 ~~,~~ \mathcal{L}_{\hat{H}} = \frac{
\hat{H}}{m_h^2} |\Box H|^2 ~~,
\ee
where $m_h$ is the physical Higgs mass. The operator $\mathcal{O}_\Box = |\Box H|^2$, where $\Box \equiv D^\mu D_\mu$, is the sole one that modifies the form of the Higgs boson propagator at dimension six. Hence a constraint on the $\hat{H}$-parameter can, in this basis, be thought of as a constraint on how the SM Higgs boson propagates.\footnote{All of these operators may be traded for different sets of operators by field redefinitions.  However, when interpreted as arising from new physics interacting with the gauge and Higgs bosons, at leading order it is instructive and convenient to work in this basis.}

The paper is organised as follows. As a prelude to our discussion, in \Sec{sec:KL} we derive general information on UV corrections to two-point functions, such as the Higgs boson self-energy, by studying the \KL~representation.  These results are employed to determine consistency conditions on the sign of the $\hat{H}$-parameter as well as the momentum expansion.  The physical interpretation of these results is also illustrated with some examples.

In \Sec{sec:theory} we discuss the EFT interpretation of $\mathcal{O}_\Box$ from a number of directions.  Our analogy begins with the precision EW parameters, which have an obvious UV interpretation in the context of scenarios in which all new physics interacts primarily with the gauge and Higgs sector, known as the `Universal' class of EFTs.  We also show that, even within the restricted class of Universal theories, the on-shell Higgs coupling measurements alone cannot unambiguously constrain the $\hat{H}$-parameter, making it a prime and challenging phenomenological target for future Higgs studies.  In \Sec{sec:UV} we then provide explicit examples of UV completions that illustrate how $\mathcal{O}_\Box$ emerges at low energy together with other operators involving the Higgs field.

In \Sec{sec:pheno} we study phenomenological aspects of $\mathcal{O}_\Box$ and show that, whenever an EFT description is valid, the commonly considered process for off-shell Higgs physics $gg\to h^\ast \to ZZ$ is in fact insensitive to the energy-growing contribution from the $\hat{H}$-parameter, making this a poor probe of off-shell Higgs behaviour in this context.  On the contrary, we demonstrate that $t \bar{t} t \bar{t}$ production provides a complementary future probe of the $\hat{H}$-parameter and off-shell Higgs physics.

\section{Prelude: \KL~and EFT}
\label{sec:KL}
%%%%%%%%%%%%%%%%%
We begin in a spirit of generality, to gain some theoretical insight on features of UV modifications of the Higgs propagator without yet committing to specific examples.  Consider the renormalised Higgs field in the broken phase.  Since it is a quantum operator, it must have a \KL~ representation and, since it is renormalised, it has a pole of unit residue at $p^2 = m_h^2$.  In momentum space the two-point function is
\be
 \Delta_{h} (p^2) = -i \int d^4 z ~e^{i p z} \langle 0 | T \{h (z) h (0) \} |  0 \rangle~.
\ee
This Green's function has a \KL~representation \cite{Kallen:1952zz,Lehmann:1954xi}, given by
\be
\Delta_{h} (p^2) =  \int^\infty_{0} dq^2 \frac{\rho_h(q^2)}{p^2-q^2+i \epsilon} ~~,
\ee
{where the spectral density function must be real and positive definite: $\rho_h(q^2)\geqslant 0$.\footnote{Note that, $\rho_h (q^2) \propto \sum_n \delta (q^2 - m^2_n) | \langle 0 | h(0)  |n \rangle |^2$, where $|n \rangle$ is a state in the Hilbert space. }
Assume that the operator $h$ has, in addition to the usual SM contributions, non-vanishing matrix elements with heavy BSM states $X$ with invariant mass above a certain mass gap $M$. This is simply the assumption that an EFT treatment below $M$ is appropriate.  Under these general conditions we can split the sum over Hilbert space,
\be
\rho_h(q^2) = \rho_{\text{SM}}(q^2) + \rho_X(q^2)~,
\ee
where $\rho_{\text{SM}}$ is the contribution to the spectral density function from the pure SM states, while the new-physics contribution is such that 
\be
\rho_X(q^2<M^2) =0~. 
\ee}
For $p^2 < M^2$ we may expand $\Delta_{h} (p^2)$ to find
\be
\Delta_{h} (p^2) =  \Delta_{\text{SM}} (p^2) - \frac{1}{M^2} \sum_{n=1}^\infty c_n \left(\frac{p^2}{M^2} \right)^{n-1}~~,
\label{eq:deltah}
\ee
where $\Delta_{\text{SM}}$ is the Higgs propagator including quantum corrections from SM degrees of freedom and
\be
c_n = M^2 \int_0^1 dx \,  \rho_X(M^2/x) \, x^{n-2} ~.
\label{eq:cn}
\ee
Thus, even though we do not know the nature of the states that the Higgs may be coupled to, we can conclude that for $p^2 \ll M^2$ all new-physics corrections to the Higgs propagator are expressed as a polynomial in $p^2/M^2$, as expected from an EFT description.

\subsection{Consistency conditions}
\label{sec:conditions}
From the result in \Eq{eq:cn} we can derive some general consistency conditions on the coefficients $c_n$ of the EFT expansion that follow from the \KL~representation.

\begin{figure}[tbp]
\begin{center}
\includegraphics[width=0.55\textwidth]{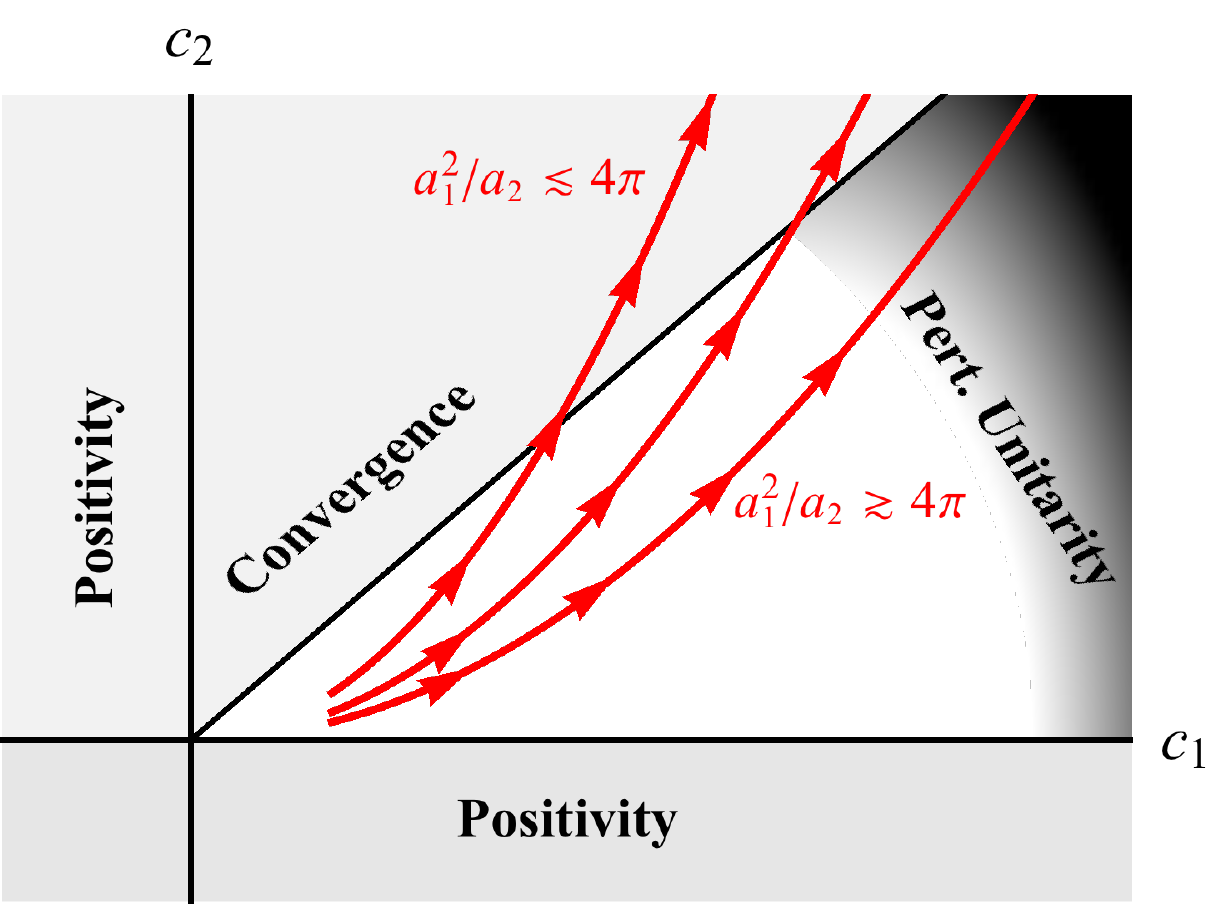}
\caption{In the plane spanned by $c_1$ and $c_2$ (the two leading coefficients of the propagator derivative expansion) we show how the constraints from {\it (i) positivity, (ii) convergence, (iii) perturbative unitarity} single out a theoretically-allowed bounded region. An experimental measurement of $a_1$ and $a_2$ (the first two terms in a momentum expansion) selects the curve $c_2 = c_1^2\, a_2/a_1^2$. Examples of these curves (for different values of $a_2/a_1^2$) are shown by solid red lines, which are generated by varying the cutoff mass $M$. The value of $M$ increases along the direction of the arrows. The stronger bound on $M$ comes from {\it convergence} when $a_1^2/a_2 \lesssim 4\pi$ and from {\it perturbative unitarity} when $a_1^2/a_2 \gtrsim 4\pi$.}
\label{fig:kl-plot}
\end{center}
\end{figure}

\subsubsection* {(i) Positivity}

We observe from \Eq{eq:cn} that the \KL~representation requires all coefficients of the EFT expansion to be positive
\be
c_n \geqslant 0 ~~~\forall n ~~~~~~~~~{\rm (positivity)}.
\label{eq:posit}
\ee
Also, either all coefficients are strictly positive ($c_n >0$ $\forall n$) or they all vanish simultaneously ($c_n =0$  $\forall n$). 

This result is reminiscent of the positivity constraints derived in \cite{Adams:2006sv}, and is relevant to our study because it implies that the Higgs oblique parameter is positive ($\hat{H} \geqslant 0$) in typical QFT UV-completions.   When applied to EW gauge bosons, the same logic implies that the oblique parameters $\hat{Y}$ and $\hat{W}$ must be positive, as observed in ref.~\cite{Cacciapaglia:2006pk}. The same authors also pointed out that if the SM gauge group is extended in the UV, then additional ghost states in the UV completion could contribute negatively to $\rho_\mathcal{O}(q^2)$, invalidating the positivity condition.  This caveat also applies for the Higgs {when the operator $h$ has matrix elements with unphysical negative-norm states.}

\subsubsection* {(ii) Convergence}

A further consequence of \Eq{eq:cn} is
\be 
c_n \geqslant c_{n+1} ~~~\forall n ~~~~~~~~~{\rm (convergence)}.
\label{eq:converg}
\ee
This inequality is saturated in the case of single-state tree-level exchange in which $\rho_X (q^2)\propto \delta (q^2-M^2)$ and all $c_n$ are equal. The condition in \Eq{eq:converg} implies that higher orders in the EFT expansion are not only suppressed by additional powers of $p^2/M^2$ (which is smaller than one, whenever the EFT is valid), but their corresponding Wilson coefficients $c_n$ also become progressively smaller. This means that the EFT series is absolutely convergent, since \Eq{eq:converg} ensures that D'Alembert's criterion is satisfied.
This is the reason for referring to this as the `convergence' condition in \Eq{eq:converg}.

The `convergence' condition becomes particularly useful when one tries to infer information on the range of validity of the EFT from the truncation of the derivative series. We will return to this important point in sect.~\ref{sec:self}. 

The `convergence' condition could be in principle checked experimentally by making precise measurements sensitive to higher-order effects in the EFT expansion. From the EFT point of view, the Wilson coefficients $c_n$ are not observables, but only the combinations $a_n \equiv c_n/M^{2n}$ are measurable. Suppose that one could measure two successive coefficients $a_n$ and $a_{n+1}$. For any set of EFT operators satisfying the \KL~representation, the `convergence' condition in \Eq{eq:converg} implies that the mass scale characterising the onset of new physics must satisfy
\be
M^2 \leqslant \frac{a_n}{a_{n+1}}  ~~~\forall n ~~~~~~~~~{\rm (convergence)}.
\label{eq:KLbound}
\ee
Thus, if consecutive powers in the EFT expansion were measured, one could in principle place a theoretical upper bound on the value of the true cutoff which, as we will show in the following, could be more restrictive than the constraint derived from requiring perturbative unitarity.

\subsubsection* {(iii) Perturbative unitarity}

An upper bound on the coefficients $c_n$ can be obtained by imposing perturbative unitarity. Consider a two-to-two scattering process mediated at tree-level by Higgs exchange. We require that the corresponding amplitude must satisfy the unitarity constraint following from the optical theorem for any energy within the validity of the EFT. In practice, this means setting $s=M^2$ in the scattering amplitude and translating the unitarity bound into a constraint on the coefficients $c_n$. 

The corresponding bound is process-dependent but roughly corresponds to a limit of order $4 \pi$ on a linear combination of the $c_n$, leading to 
\be 
c_n \lesssim 4 \pi ~~~\forall n ~~~~~~~~~{\rm (perturbative~unitarity)}.
\ee
A precise determination of the limit is not possible, since the choice $s=M^2$ means that we are working at the edge of the EFT validity and the expansion is not under control. 

\subsubsection* {Combining the three conditions}

It is interesting to compare the impact of the three conditions (`positivity', `convergence', `perturbative unitarity') on the allowed values of the Wilson coefficients. This can be simply done by restricting our considerations to the first two coefficients in the EFT expansion in \Eq{eq:expansion} and visualising the conditions in the plane $c_1$--$c_2$, as shown in~\Fig{fig:kl-plot}. This figure illustrates the complementarity of the different conditions which, when combined, single out a special region which is the only one allowed by theoretical considerations.

Experiments cannot directly determine $c_{1,2}$ but measurements of $a_{1,2}$ identify a curve in the plane of~\Fig{fig:kl-plot}. Varying the unknown cutoff $M$ will trace out the parabola $c_2 = c_1^2 (a_2/a_1^2)$. This curve starts at the finite value $c_1 = a_1 E^2$, where $E$ is the typical energy of the process at which $a_1$ is measured.\footnote{{The SM radiative corrections could imply some residual soft dependence on the scale $M$.}} Lower values of $c_1$ violate the EFT validity.

As we increase the value of $M$, we move up along the curve until we hit either the `convergence' or the `perturbative unitarity' bound. This establishes a limit on the new-physics mass $M$. Whenever $a_1^2/a_2\lesssim 4\pi$, `convergence' gives a stronger limit on $M$ than the more familiar `perturbative unitarity' limit, see~\Fig{fig:kl-plot}. 

\subsection{From propagator to self-energy}
\label{sec:self}

For practical calculations of low-energy effects from new physics, one starts from the self-energy $\Sigma_h$ rather than the propagator $\Delta_h$. The translation -- at the non-perturbative level -- can be made through the Dyson equation $\Delta_h = \Delta_{\text{SM}} (1+\Sigma_h \Delta_h)$, which gives
\be
\Sigma_h (p^2) = \Delta_{\text{SM}}^{-1} (p^2) - \Delta_h^{-1} (p^2) ~ .
\label{eq:dyson}
\ee
Using the expansion in \Eq{eq:deltah} and taking for simplicity $\Delta_{\text{SM}}^{-1}=p^2-m_h^2$, we find the EFT expansion for the self-energy
\be
\Sigma_h (p^2) = - (p^2 -m_h^2)  \sum_{n=1}^\infty {\hat c}_n \left(\frac{p^2}{M^2} \right)^{n}~,
\label{eq:sigm}
\ee
\be
{\hat c}_n = \left(1- \frac{m_h^2}{p^2} \right) \left( c_n + \sum_{j=1}^{n-1} c_j \, {\hat c}_{n-j} \right) ~.
\label{eq:hatcn}
\ee
In the following, for simplicity, we consider the case $p^2 \gg m_h^2$ and set $m_h=0$.

From the recursive relation in \Eq{eq:hatcn}, we infer several properties of the Wilson coefficients ${\hat c}_{n}$. First, from the positivity of $c_n$ we conclude that all ${\hat c}_{n}$ are positive as well. Second, ${\hat c}_1 = c_1$ and ${\hat c}_n \geqslant c_n(1+c_n)^{n-1}$ for $n>1$, with the inequality being saturated for single-particle tree-level exchange (corresponding to $c_n$ all equal for any $n$). Third, contrary to $c_n$ which satisfy the `convergence' condition, the coefficients ${\hat c}_n$ can grow with $n$ and diverge. In particular, the progression of ${\hat c}_n$ diverges (strictly violating the `convergence' criterion) if any of these conditions is satisfied\footnote{These results follow directly from the definition of ${\hat c}_n$. Indeed, from \Eq{eq:hatcn} we obtain ${\hat c}_{n+1} -c_1 {\hat c}_{n}= c_{n+1} + \sum_{j=1}^{n-1} c_{j+1} {\hat c}_{n-j} \geqslant 0$. Hence, we derive condition ({\it i}). Next, consider the inequality
${\hat c}_n \geqslant  c_n(1+c_n)^{n-1}$. If either condition ({\it ii}) or ({\it iii}) is verified, then the right-hand side diverges for $n \to \infty$;  hence the progression of ${\hat c}_n$ diverges as well. If conditions ({\it i})--({\it iii}) are not verified, ${\hat c}_n$ do not necessarily diverge. For instance, taking $c_n = c_1/n^\alpha$, the progression of ${\hat c}_n$ remains finite whenever $c_1 <1$ and $\alpha > \alpha_c$, where $\alpha_c$ starts at 1/2 for small $c_1$ and grows with $c_1$.}: ({\it i}) $c_1>1$;  ({\it ii}) $\lim_{n\to \infty} c_n \ne 0$; ({\it iii}) $c_n$ approaches zero at large $n$ slower than $c_n \sim n^{-1/2}$. 

The property of `convergence' guarantees that one can consistently extract information about the validity range of the EFT from a truncation of the perturbative series, with a precision that grows with the number of retained terms. On the contrary, this cannot be done reliably whenever `convergence' is not satisfied (as in the case of the derivative expansion of the self-energy with $c_1>1$) because higher-order terms neglected in the truncation can be larger than the terms retained. 

As an example of this problem, consider a truncation of the self-energy expansion in \Eq{eq:sigm}, keeping only the four-derivative term corresponding to $n=1$. This predicts a ghost with mass $M/\sqrt{c_1}$. If $c_1<1$, the ghost lies above the EFT cutoff. If $c_1>1$, the ghost is below the cutoff, indicating a premature breakdown of the momentum expansion at energies below the true cutoff $M$. However, this prediction is unreliable since the progression of ${\hat c}_n$ diverges (for $c_1>1$) and the conclusion is based on a truncation in which the terms neglected are larger than those retained. 
To find the correct answer, we must turn to the derivative expansion of the propagator in \Eq{eq:deltah}, which is always under control as it satisfies `convergence'. From this expansion, we do not find any ghost: $M/\sqrt{c_1}$ is the energy at which new-physics effects become larger than the SM contribution, but the derivative expansion breaks down only at the scale $M$. The `convergence' condition $c_{n+1} < c_n$ maintains validity of the momentum expansion in the amplitude until $p^2 = M^2$.

In conclusion, the correct recipe is to expand the full propagator in powers of $p^2/M^2$, rather than keeping the expansion of the self-energy in the denominator of the propagator. `Convergence' insures the correctness of the EFT interpretation of the results based on this recipe even when $c_1 >1$, since higher-order terms in the derivative expansion are consistently smaller, all the way up to the physical cutoff.  In \Sec{app:toyexample} we will provide an explicit extra-dimensional example where this becomes particularly apparent.

\begin{figure}[tbp]
\begin{center}
\includegraphics[width=0.3\textwidth]{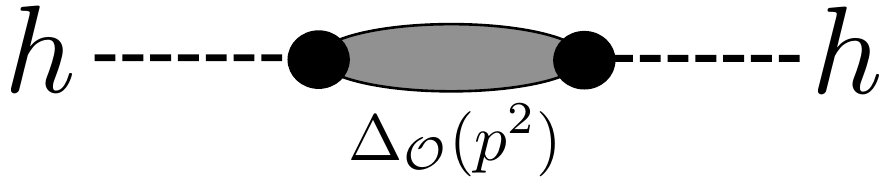}
\caption{Higgs self-energy correction from the two-point function of the operator $\mathcal{O}$.}
\label{fig:diagrams}
\end{center}
\end{figure}

\subsection{Perturbative perspective}
More practically, to compute new-physics effects, one often performs a perturbative calculation, relying on the assumption that the new-physics sector is weakly coupled to the Higgs. To this end, consider a generic interaction of the bare Higgs $h^0$ with other new BSM fields
\be
\mathcal{L} = \mathcal{L}_{\text{SM}} + \mathcal{L}_{\text{int}}  ~~,
\ee
where in $\mathcal{L}_{\text{int}}$ we couple the Higgs boson to some additional external operator $\mathcal{O}$ as
\be
\mathcal{L}_{\text{int}} = \kappa \, h^0 \mathcal{O}  ~~.
\label{eq:cappa}
\ee
We take the coupling constant $\kappa$ to be dimensionless and absorb all dimensionful parameters in the definition of $\mathcal{O}$.  The correction to the Higgs boson self-energy,
\be
\Sigma^0_h (p^2) = \kappa^2 \Delta_{\mathcal{O}} (p^2) = - i \kappa^2 \int d^4 z ~e^{i p z} \langle 0 | T \{\mathcal{O} (z) \mathcal{O} (0) \} |  0 \rangle  ~~,
\ee
occurs at $\mathcal{O}(\kappa^2)$ (see \Fig{fig:diagrams}). The self-energy correction is related to the two-point function for $\mathcal{O}$, which also has a \KL~representation, given by
\be
\Delta_{\mathcal{O}} (p^2) =  \int^\infty_{0} dq^2 \frac{\rho_\mathcal{O}(q^2)}{p^2-q^2+i \epsilon} ~~.
\ee
Now one can proceed with a derivative expansion, identical to the analysis performed at the beginning of this section for $\Delta_h$.

Depending on the form of the operator $\mathcal{O}$, the lowest order terms in the $p^2$ expansion of $\Sigma^0_h (p^2)$ may not be finite.  However, we will assume that the underlying theory is renormalisable, such that only $\Sigma^0_h (0)$ and $d\Sigma^0_h (p^2)/dp^2 \big|_{p^2=0}$ contain divergences which are absorbed by mass and wavefunction renormalisation for the Higgs.  This restricts the set of UV theories under consideration because, unlike the case of $2\to2$ $S$-matrix amplitudes, here we have no strict upper bound on the number of subtractions that may be required in a general quantum field theory (QFT). Put another way, for two-point Green's functions we do not have a constraint analogous to the Froissart bound \cite{Froissart:1961ux}.   Furthermore, even for $2\to2$ $S$-matrix elements where one can apply the Froissart bound it is not, in full generality, possible to rule out the requirement for a subtraction which corresponds to a dimension-6 operator in the EFT.  As a result, typically only dimension-8 EFT operators can be constrained with analyticity arguments.  Nonetheless, assuming less generality, it is still possible to set bounds on dimension-6 operators, as was considered in \cite{Bellazzini:2014waa,Low:2009di,Falkowski:2012vh,Urbano:2013aoa}.\footnote{For further related discussion see \cite{Bellazzini:2016xrt}.} Here by assuming that mass and wavefunction counterterms suffice, as this hypothesis applies to the renormalisable QFTs we typically encounter in weak-scale models, the scope of applicability is limited to specific classes of UV theories.

Nonetheless, proceeding with this assumption, we write the renormalised self-energy as
\be
\Sigma_h (p^2) = \Sigma^0_h (p^2) - \delta_m -  (p^2-m_h^2)\, \delta_p ~~,
\ee
and choose to canonically normalise the Higgs field and set its mass to the physical value through the choice
\be
\delta_m = \Sigma^0_h (m_h^2) ~,~~~~~~ \delta_p = \frac{d\Sigma^0_h (p^2)}{d p^2} \bigg|_{p^2=m_h^2}  ~~.
\label{eq:CTs}
\ee 
Including the mass gap and the renormalisation conditions, the renormalised self-energy takes the twice-subtracted form
\be
\Sigma_h (p^2) = \kappa^2  \int^\infty_{M^2} dq^2 \left(\frac{p^2-m_h^2}{q^2-m_h^2}\right)^2 \frac{\rho_\mathcal{O}(q^2)}{p^2-q^2+i \epsilon}  ~~.
\ee

For $p^2 \ll M^2$, which is the case of interest for EFT considerations, we have that $\Sigma_h (p^2)$ is real.  By Taylor expanding we find
\be
\Sigma_h (p^2)  = - \left( p^2-m_h^2\right)\, \sum_{n=1}^\infty \, C_n \, \left( \frac{p^2}{M^2}\right)^n \, f_n\left( \frac{m_h^2}{p^2}\right) ~,
\label{eq:expansion}
\ee
\be
C_n = \kappa^2 \int_0^1 dx \, \frac{ \rho_\mathcal{O}(M^2/x)}{M^2} \, x^n ~,
\ee
\be
f_n(y) =\frac{1-(n+1)y^n+ny^{n+1}}{1-y} ~.
\label{eq:fn}
\ee
Since they are obtained from a \KL~decomposition, the coefficients $C_n$ satisfy the same conditions of positivity and convergence that were derived for the coefficients $c_n$ in \Sec{sec:conditions}. {Note that, for $p^2\gg m_h^2$ and at $\mathcal{O}(\kappa^2)$, $c_n = \hat c_n = C_n$.
To conclude, at leading order in $\kappa$, both the propagator and the self-energy expansion in $p^2 / M^2$ obey convergence criterion.}

These observations are made with a view towards practical calculations of the Higgs boson two-point function, which concerns the rest of this paper. However, the discussion in terms of the \KL~representation for composite operators opens the door to extending these results beyond two-point Green's functions.
 In particular, it may be possible to derive similar convergence conditions for the case of forward scattering amplitudes, since they have dispersion relations, somewhat analogous to that of \KL, where positivity follows from the optical theorem.  This would be advantageous as it would elevate the convergence relations to the level of scattering amplitudes, eliminating the need for any consideration of EFT bases.  For illustration, in \App{app:fun} we include a jovial application of convergence to string theory amplitudes.

\subsection{Scherzando: gedanken measurements}
\label{app:muondecay}

In this spirit we will present some examples of how experimental measurements combined with the `convergence' condition can lead to stringent constraints on the cutoff mass $M$, derived from a purely low-energy perspective. Although these example are fictitious, as they are based on EFT of which we already know the UV completion, they illustrate the procedure that can be in principle applied to future experiments where the SM plays the role of the EFT. These examples also explain how the `convergence' condition can be of utility in scenarios that go beyond two-point functions.  

\subsubsection*{Muon decay}
As a purely academic (albeit hopefully instructive) exercise, imagine a civilisation that has never performed experiments at energy higher than a few hundred MeV and instead measured muon decay ad nauseam. With impressive theoretical insight, the physicists of this unlucky civilisation assume that muon decay is mediated by a charged vector operator involving unknown UV dynamics that couples to leptons as
\be
\mathcal{L}_{\mathcal{O}^\mu} =  J_e^\alpha \mathcal{O}_\alpha +  J_\mu^\alpha \mathcal{O}_\alpha  + {\rm h.c.}~~,
\ee
where
\be
J_e^\alpha = \overline{e} \gamma^\alpha (1-\gamma_5) \nu_e ~~~,~~~ J_\mu^\alpha = \overline{\mu} \gamma^\alpha (1-\gamma_5) \nu_\mu ~~.
\ee
One can integrate out this operator using the \KL~prescription, which for a vector operator gives
\be
\Delta^{\alpha\beta}_{\mathcal{O}} (p^2) =  \int^\infty_{M^2} dq^2 \frac{g^{\alpha \beta} \rho^T_\mathcal{O}(q^2) - \rho^V_\mathcal{O}(q^2) \frac{p^\alpha p^\beta}{q^2}}{p^2-q^2+i \epsilon} ~~.
\ee

When calculating the matrix element for $\mu \to e \nu_\mu {\bar \nu}_e$, mediated by this propagator, the $p^\alpha p^\beta$ terms will generate powers of $m_e$ which can be ignored since $m_e \ll m_\mu$.  As a result, the two leading terms of the generated tower of higher-dimension operators are
\be
\mathcal{L}_{\text{EFT}} = J_e^{\alpha \dagger} \left(-a_1 + a_2\, \partial^2 \right) J_{\mu\alpha} + {\rm h.c.} ~~.
\ee

While terrestrial physicists have the privilege of knowing the SM result $a_1=g^2/8m_W^2$, $a_2=g^2/8m_W^4$, our fictitious physicists can only make the following inference from the `positivity' and `convergence' criteria in \Eq{eq:posit} and (\ref{eq:KLbound})
\be
a_1 \geqslant 0 ~, ~~~~a_2 \geqslant 0 ~,~~~~m_W^2 \leqslant \frac{a_1}{a_2} ~,
\label{eq:slant}
\ee
where $m_W$ is the cutoff mass.
However, our gedanken civilisation can benefit from precise experimental measurements of the differential muon decay rate, which is given by 
\be
\frac{d \Gamma_\mu}{dx} =  \frac{a_1^2\, m^5_\mu\, x^2}{48 \pi^3}  \left[ 3-2 x+   x (2-x)m_\mu^2\, \frac{a_2}{a_1} \right] ~~,
\label{eq:gammu}
\ee
where $x=2E_e/m_\mu$ with $E_e$ being the electron momentum in the LAB frame and the electron mass has been neglected. 

Suppose one had a measurement of the decay rate with a fractional uncertainty which is about a factor $6$ stronger than what is known today. Then, by binning in the final state electron energy, one could extract a $90\%$ CL lower bound on  $m_\mu^2 a_2/a_1$ at the level of $3\times 10^{-7}$. Using the `convergence' criterion in \Eq{eq:slant}, one derives a theoretical upper bound on the EFT cutoff of $m_W \lesssim 190 \text{ GeV}$.

Note that this constraint on $m_W$ is much stronger than the bound from `perturbative unitarity' of the Fermi theory ($m_W \lesssim \sqrt{4\pi} \, v \sim 900$~GeV), as it could have been guessed from the start since the condition $a_1^2/a_2 \lesssim 4\pi$ is amply satisfied in the SM. 

Here, for simplicity of presentation, we have neglected mass corrections $O(m_e^2/m_\mu^2)$ and radiative corrections $O(\alpha /\pi)$, but these can be included in a more realistic calculation of the bound on $m_W$. However, it is important to stress that none of these IR effects can generate $O(E_e^4)$ terms in $d \Gamma_\mu /dE_e$, which are instead induced by $a_2$, see \Eq{eq:gammu}. These energy-growing terms are characteristic of $a_2$ and are the reason for the enhanced sensitivity on the UV features of the theory.

By improving further the precision on the measurements of the muon decay energy spectrum and the EFT theoretical prediction by computing QED radiative corrections up to the appropriate loop order, one could obtain tighter bounds from `convergence', in principle all the way up to saturating the physical value of $m_W$. This example shows how the `convergence' criterion combined with precise measurements can yield information about the EFT cutoff mass.

\subsubsection*{Lepton forward-backward asymmetry}

Imagine now a slightly more advanced civilisation that can build high-energy colliders, although without reaching the threshold for weak gauge boson production. Those physicists can measure the forward-backward asymmetry in $e^+ e^- \to \mu^+ \mu^-$, {\it i.e.} the normalised difference in the number of events in the forward and backward hemispheres as defined by the same-charge flow. At energies below the $Z$-boson resonance, the effect comes from the interference between photon exchange and an axial-vector four-fermion interaction parametrised as
\be
\mathcal{L} = (\bar e \gamma_\mu \gamma^5 e) \left(-a_1 + a_2\, \partial^2 \right) (\bar \mu \gamma^\mu \gamma^5 \mu) ~,
\label{eq:LagWET}
\ee
truncating the expansion at dimension-8. In the SM at leading order, $a_1 = G_{F}/{2\sqrt{2}}$, while $a_1 / a_2 = m_Z^2$. In the EFT, the forward-backward asymmetry is given by
\be
A_{FB} (s) = - \frac{3\,a_1 \, s \, \left(1+\frac{a_2}{a_1} \, s\right)}{8\, \pi \, \alpha^2}~,
\ee
where $\sqrt{s}$ is the centre-of-mass energy and $\alpha$ is the QED structure constant.  The term proportional to $a_2 / a_1$ grows with the collider energy. 

Just as an example, we fit all available PDG data~\cite{Tanabashi:2018oca} on $e^+e^-$ in the range $\sqrt{s} = 29$--$45$~GeV. Profiling over $a_1$, we find $a_2 / a_1 > (170$~GeV$)^{-2}$ at $90\%$ CL which, using the `convergence' condition in \Eq{eq:KLbound}, translates into the bound $m_Z \lesssim 170\,\text{ GeV}$.  This example, when compared to the case of muon decay, shows the importance of probing the EFT at higher energy. Since one is after the term $E^2 a_2/a_1$, where $E$ is the typical energy of the process, similar bounds on the cutoff mass $M$ can be obtained with limited precision at high energy or with high precision at low energy.

We conclude this section by recalling the academic spirit of our discussion. The application of this procedure is practically limited by the fact that other unknown new-physics effects make the extraction of the propagator corrections in general ambiguous.  Closing this digression, we return to the case at hand, which is the SM.

\section{Universal EFTs}
\label{sec:theory}

\subsection{Operator analysis}
Before considering the general phenomenological picture for $\mathcal{O}_\Box$, we will discuss the broader context into which this operator fits.  Looking at the microscopic origin of dimension-6 operators in the EFT, save for one specific example we will return to later, we expect that general new physics scenarios will not generate only the operator $\mathcal{O}_\Box$ at the matching scale, but also a variety of other operators. 

\begin{table}[tt]
\centering

{\bf `Higgs-only'} \\ \medskip 
\begin{tabular}{c c c}
{\boldmath$[g_\ast^0]$} & {\boldmath$[g_\ast^2]$} & {\boldmath$[g_\ast^4]$} \\
$\mathcal{O}_\Box = \frac{c_\Box}{M^2} |\Box H|^2 $  \qquad &  \qquad $\mathcal{O}_H =   \frac{c_H}{2M^2}  \left(\partial^\mu |H|^2 \right)^2 $  \qquad &  \qquad $\mathcal{O}_6 =  \frac{c_6}{M^2}  |H|^6 $  \\
$$  \qquad & \qquad$\mathcal{O}_T =  \frac{c_T}{2M^2}  (H^\dagger {\overleftrightarrow D}^\mu H)^2 $  \qquad &  \qquad $$\\
 $$  \qquad &  \qquad $\mathcal{O}_R = \frac{c_R}{M^2} |H|^2 |D^\mu H|^2 $  \qquad &  \qquad $$   \\ [1ex] \hline \\
\end{tabular}

{\bf `Gauge-only'} \\ \medskip
\begin{tabular}{c c c}
$\mathcal{O}_{2G} = -\frac{c_{2G}}{4 M^2} (D_\rho G^a_{\mu\nu})^2 $  \qquad& \qquad 
$\mathcal{O}_{2W} = -\frac{c_{2W}}{4 M^2} (D_\rho W^a_{\mu\nu})^2 $  \qquad&  \qquad
 $\mathcal{O}_{2B} = -\frac{c_{2B}}{4 M^2} (\partial_\rho B_{\mu\nu})^2 $    \\ [1ex] \hline \\
\end{tabular}

{\bf `Mixed gauge-Higgs'} \\ \medskip
\begin{tabular}{c c}
$\mathcal{O}_B =  \frac{i g'\, c_B}{2M^2} (H^\dagger  {\overleftrightarrow D}^\mu H) \partial^\nu B_{\mu\nu} $ \qquad & \qquad $\mathcal{O}_{GG} =   \frac{g_s^2\, c_{GG}}{M^2} |H|^2 G^{a,\mu\nu} G^a_{\mu\nu} $  \\
$\mathcal{O}_W =  \frac{i g \, c_W}{2M^2} (H^\dagger \sigma^a {\overleftrightarrow D}^\mu H) D^\nu W^a_{\mu\nu} $ \qquad & \qquad$\mathcal{O}_{WB} = \frac{g g' \, c_{WB}}{M^2} H^\dagger \sigma^a H B^{\mu\nu} W^a_{\mu\nu} $ \\
$$ \qquad & \qquad $\mathcal{O}_{WW} =   \frac{g^2\, c_{WW}}{M^2} |H|^2 W^{a\,\mu\nu} W^a_{\mu\nu} $   \\
$$ \qquad & \qquad $\mathcal{O}_{BB} =   \frac{g'^2\, c_{BB}}{M^2} |H|^2 B^{\mu\nu} B_{\mu\nu} $  \\ [1ex] \hline \hline \\
\end{tabular}

{\bf Relations between oblique parameters and Wilson coefficients} \\ \medskip
\begin{tabular}{c c}
$\hat{S} =4 \left(c_{WB}+\frac{c_W+c_B}4 \right) \frac{m_W^2}{M^2}$ \qquad & \qquad  
$\hat{T} = c_T\, \frac{v^2}{M^2}$ \\
$\hat{W} =c_{2W}\, \frac{m_W^2}{M^2}$ \qquad & \qquad
$\hat{Y} =c_{2B}\, \frac{m_W^2}{M^2}$ \\
$\hat{Z} =c_{2G}\, \frac{m_W^2}{M^2}$ \qquad & \qquad
$\hat{H} = c_\Box \, \frac{m_h^2}{M^2}$
\\ [1ex] \hline \\
\end{tabular}

\caption{The complete set of CP-even operators (up to total derivatives) in the Universal basis, as they appear in the Lagrangian, divided into three classes: `Higgs-only' (operators containing only the Higgs doublet and covariant derivatives), `gauge-only' (operators containing gauge field strengths and covariant derivatives), and `mixed gauge-Higgs'. The Wilson coefficients of `Higgs-only' operators carry the power of the Higgs sector couplings (generically denoted by $g_\ast$) as indicated in the table. The Wilson coefficients of `gauge-only' and `mixed gauge-Higgs' operators are dimensionless (in units of coupling). We also give the relations between oblique parameters and Wilson coefficients, which are valid in the Universal basis. We have chosen $v\approx246$~GeV. }
\label{table:Mixed}
\end{table}

With this in mind, there is a very broad class of UV theories which single out a particular set of EFT operators at the matching scale, within which the $\hat{H}$-parameter is well defined as the Wilson coefficient of $\mathcal{O}_\Box$.   This is none other than the class of Universal theories~\cite{Barbieri:2004qk,Wells:2015uba}.  Here we broadly define an EFT to be Universal when there exists a field basis in which all leading-order effects are captured at dimension 6 by operators containing only SM bosonic fields. The complete list of these operators (up to total derivatives) is given in \Tab{table:Mixed}.  Note that this definition captures all scenarios in which new heavy states interact primarily with the bosons of the SM.  It also captures scenarios in which the new physics couples to quarks and leptons through the SM gauge currents $J_W^\mu$, $J_B^\mu$ and $J_G^\mu$, or to the SM Higgs scalar current $J_H$, which we define as
\be
 J_H =  \mu^2 H - 2 \lambda |H|^2 H - \bar q i\sigma_2 Y^\dagger_u u - \bar d Y_d q - \bar e Y_e \ell~ .
 \label{eq:JH}
\ee
This is because, through appropriate field redefinitions, the generated operators involving these currents can be rewritten in terms of bosonic fields only.  Similarly, operators containing quarks and leptons in exactly the same combination as the SM scalar current can be redefined by using the Higgs equation of motion $\Box H = J_H$.

In many conventional EFT bases~\cite{Giudice:2007fh,Grzadkowski:2010es,Elias-Miro:2013eta}, for computational convenience the operator $\mathcal{O}_\Box$ is replaced with $J_H^2$ after field redefinition. Here, we prefer to work in a `boson-only' basis, which more clearly matches with the UV properties of a Universal theory where new physics is coupled only to EW and Higgs bosons. 

In \Tab{table:Mixed}, we have separated the Universal operators into three classes: `Higgs-only', `gauge-only', and `mixed gauge-Higgs'. The `Higgs-only' operators have been ordered according to their dimension in units of coupling constant (for notation, see sect.~2.1 of ref.~\cite{Giudice:2016yja}). Note that the ordering in terms of coupling dimension is useful in charting the space of microscopic completions.  For instance, $\mathcal{O}_\Box$ and $\mathcal{O}_6$ lie at two extremes of the coupling spectrum.  Since the Wilson coefficient for $\mathcal{O}_6$ is $\mathcal{O}(g_\ast^4)$, it will typically be large in strongly coupled completions, but small in weakly coupled completions.  On the other hand the Wilson coefficient for $\mathcal{O}_\Box$ may survive even in very weakly coupled completions.  These extremes, and the territory in between, will be discussed in \Sec{sec:UV} in some specific examples of UV completions.

Although covering an interesting and broad class of models, Universal EFTs do not match to all microscopic theories. {Moreover, the Universal basis is not closed under quantum corrections, i.e. the RG evolution~\cite{Jenkins:2013zja,Jenkins:2013wua,Alonso:2013hga} from the matching scale to the IR scale will typically populate operators not contained in the Universal basis~\cite{Wells:2015cre}.} Hence, next-to-leading order effects due to degrees of freedom both within and beyond the SM are not, in general, captured by an analysis limited to operators in the Universal basis.

\subsection{Physical effects}
\label{sec:physeff}
The most characteristic effect of the oblique parameter $\hat{H}$ (in the Universal basis) is a modification of the SM Higgs boson propagator which, for a canonically normalised field and after mass redefinition, is
\be
\Delta_h(p^2) =  \frac{1}{p^2-m_h^2} -\frac{ \hat{H}}{m_h^2} \, .
\label{eq:propagator}
\ee
Note that it is important to expand the propagator to dimension-6 here since, as discussed in \Sec{sec:self}, when the Wilson coefficients are large the dimension-8 terms in the self-energy may play an important role in cancelling the squared dimension-6 contribution.

We see the direct analogy with the definition of the EW oblique parameters $\hat{W}$ and $\hat{Y}$ through the relation with the Higgs self-energy
\be
\hat{H} = -\frac{m_h^2}{2} \Sigma_h^{\prime\prime} (m_h^2)  ~~.
\label{eq:3.3}
\ee
Thus we interpret the $\hat{H}$ parameter as sourcing a modification of the Higgs propagator which, as shown in \Eq{eq:propagator}, corresponds to a new contact term. This interpretation is of course basis-dependent, much like, for example, the value of the Higgs quartic coupling is basis dependent in an EFT.  However, within Universal UV completions, this modified-propagator interpretation is of utility.

In addition to the propagator correction, Higgs couplings are also modified.  In this section, for illustration purposes, we will focus on the effect of `Higgs-only' operators. In this regard, the interaction between a single Higgs and two gauge bosons is modified with respect to the SM couplings as follows
\be
\mathcal L = \left( g^{\rm SM}_{hWW} \, W^{+\mu}W^-_\mu \, {\mathcal D}_W +g^{\rm SM}_{hZZ} \, Z^{\mu}Z_\mu \, {\mathcal D}_Z \right)  h \, ,
\ee
\be
{\mathcal D}_W =1 + (c_R-c_H)\frac{v^2}{2 M^2} -\hat{H} \, \left( 1 + \frac{\partial^2}{m_h^2} \right) \, ,
~~~~~{\mathcal D}_Z = {\mathcal D}_W -2 \hat{T}
 \, ,
 \label{eq:gaugeH}
\ee
where $v \approx 246$~GeV.
This result has been obtained by taking into account both the Higgs wave-function rescaling and the modification of the SM relation between $v$ and $m_W$ due to Universal `Higgs-only' operators. Note that, because of the strong experimental constraints on violations of custodial symmetry in EW data, the difference between ${\mathcal D}_Z$ and ${\mathcal D}_W$ is negligible for the precision that can be achieved in Higgs physics. Thus, the modification of Higgs couplings to gauge bosons is practically identical for $W$ and $Z$.

As apparent from \Eq{eq:gaugeH}, 
the $\hat{H}$-dependent correction to the coupling with gauge bosons vanishes for on-shell Higgs bosons, where $(\partial^2  + m_h^2)h=0$. It vanishes for off-shell Higgs as well, since the corrections to the propagators and vertex exactly cancel out, at the order in which we are working:
\be
\left( \frac{1}{p^2-m_h^2} -\frac{ \hat{H}}{m_h^2} \right)
\left[ 1 -\hat{H} \, \left( 1 - \frac{p^2}{m_h^2} \right) \right]
 = \frac{1}{p^2-m_h^2} ~~.
\ee 
This result simply reflects the fact that $\mathcal{O}_\Box$ modifies the propagator for $H$ in the unbroken phase, where covariant derivatives include gauge fields. Thus the correlation between the effects in the gauge coupling and the propagator in the broken phase is a consequence of gauge symmetry.

A more direct way of understanding this cancellation comes from making a change of basis through the substitution $\Box H \to J_H$ in $\mathcal{O}_\Box$. As a result, only Higgs couplings to fermions and self-couplings show new-physics modifications, while  the Higgs-gauge coupling or multi-gauge interactions remain SM-like {(see also~\cite{Brivio:2014pfa}).}

An important consequence of this fact is that the one-loop process involving an off-shell Higgs boson, $gg \to h^\star \to ZZ$, is insensitive to modifications of the Higgs boson propagator within an EFT, since all dimension-6 terms cancel, leaving only the modification of the Higgs Yukawa coupling to the top quark which is, in any case, better constrained from on-shell measurements~\cite{Buschmann:2014sia,Corbett:2015ksa,Englert:2017aqb}.

Moving now to consider fermions, we find a universal modification of the Higgs couplings to quarks and leptons of the form
\be
\frac{y_f}{y_f^{\rm SM}} =  1 -\hat{H} - c_H \frac{v^2}{2 M^2} \, .
\label{eq:yukyuk}
\ee
In the Universal basis, this effect comes purely from the canonical rescaling of the Higgs field and the proper redefinition of $m_W$ that enters the normalisation of the SM coupling $y_f^{\rm SM}$. 

Finally, the Higgs trilinear self-coupling is modified as
\be
\frac{A_h}{A_h^{\rm SM}} =  1 -2 \hat{H} - \left(c_R+3 c_H+4 c_6 \frac{v^2}{m_h^2}\right) \frac{v^2}{2 M^2}  \, .
\ee

In conclusion, the `Higgs-only' basis is described by 4 independent Wilson coefficients ($c_\Box,c_H,c_R,c_6$) and leads to 3 physical observables in Higgs couplings: universal modifications of $h\to VV$ and $h \to \bar f f$, and the Higgs trilinear vertex. Therefore, even in this restrictive class of EFT, it is not possible to unambiguously determine $\hat{H}$ by combining on-shell Higgs coupling measurements and a measurement of the trilinear coupling.

Including the `mixed gauge-Higgs' operators adds new physical effects ($h\to gg$, $h\to \gamma \gamma$, $h\to Z \gamma$, new Lorentz structures in $h\to VV$) but also introduces several new free parameters.\footnote{For a discussion of the connection between the corrections to $h\to \gamma \gamma$ and the Higgs self-energy see \cite{Gori:2013mia}.} The only way to break the degeneracy afflicting Higgs coupling measurements is to consider alternative probes. This is because the hallmark of the $\hat{H}$ oblique parameter is off-shell Higgs physics. This strategy for unambiguously determining $\hat{H}$ at high-energy colliders will be discussed extensively in \Sec{sec:pheno}.

\section{Connecting the EFT with the UV}
\label{sec:UV}

\subsection{UV completions}
Universal EFTs describe a sm{\"o}rg{\aa}sbord of microscopic models.  Explicit calculations of the leading order Wilson coefficients for specific scenarios can be found in \cite{Henning:2014gca,Henning:2014wua,Huo:2015exa}.  Some of the examples that populate a large number of Universal operators at the same loop order, including $\mathcal{O}_\Box$, are stops in supersymmetry \cite{Henning:2014gca}, and scenarios with vector-like leptons \cite{Huo:2015exa}. 

\subsubsection*{An extra-dimensional example}
\label{app:toyexample}
Let us consider a simple extra-dimensional toy model and for simplicity take the Higgs mass to be vanishing.  This example reveals two key features.  The first is that $\hat{H}$ can be parametrically enhanced relative to $\hat{W}$ and $\hat{Y}$ in concrete extra-dimensional scenarios.  The second is that this simple example illustrates the importance of expanding the propagator consistently order-by-order in the EFT.

We take the Higgs and gauge bosons to propagate in the bulk, with the fermions localised at one end of the extra dimension.  For the Higgs we allow a bulk mass $M_{\text{Bulk}}$ and boundary conditions allowing for a massless zero mode localised away from the matter brane, the mass spectrum is $M_n^2 = M_{\text{Bulk}}^2 + n^2/R^2$.  Denoting the scale at which the EFT breaks down as $M = M_1$, the bulk mass as $M_{\text{Bulk}} = \alpha M$, and writing $p^2 = x M^2$, then we have that the full effective Higgs propagator is given by
\be
\Delta (x) = \frac{1}{2 \alpha x M^2} \left[1-\exp\left( {\frac{2 \pi \alpha}{\sqrt{1-\alpha^2}}} \right) \right] \left[ \alpha - \sqrt{x-\alpha^2} \cot \left( \pi \sqrt{\frac{x-\alpha^2}{1-\alpha^2}} \right) \right] ~~.
\ee
Expanding to dimension-6 one has
\be
c_1(\alpha) = \hat{c}_1(\alpha) = \frac{1}{4 \alpha^2} \left[ 1+\coth \left( \frac{\alpha \pi}{1-\alpha^2} \right) \right] \left[ \sinh \left( \frac{2 \pi \alpha}{1-\alpha^2} \right)  -  \frac{2 \pi \alpha}{\sqrt{1-\alpha^2}} \right] ~~,
\ee
which, due to the exponential behaviour, can be arbitrarily large, saturating even the perturbativity bound $c_1 \approx 4 \pi$ for a reasonably small bulk mass $\alpha \approx 0.35$.  For this example, since the bulk gauge boson masses are vanishing by gauge invariance, thus we have
\be
\frac{\hat{H}}{\hat{W}} = c_1(\alpha) \left(1-\alpha^2 \right) \frac{3}{\pi^2} \frac{m_h^2}{m_W^2}  ~~,
\ee
which can be arbitrarily large in this class of model.

\begin{figure}[tbp]
\begin{center}
\includegraphics[width=0.65\textwidth]{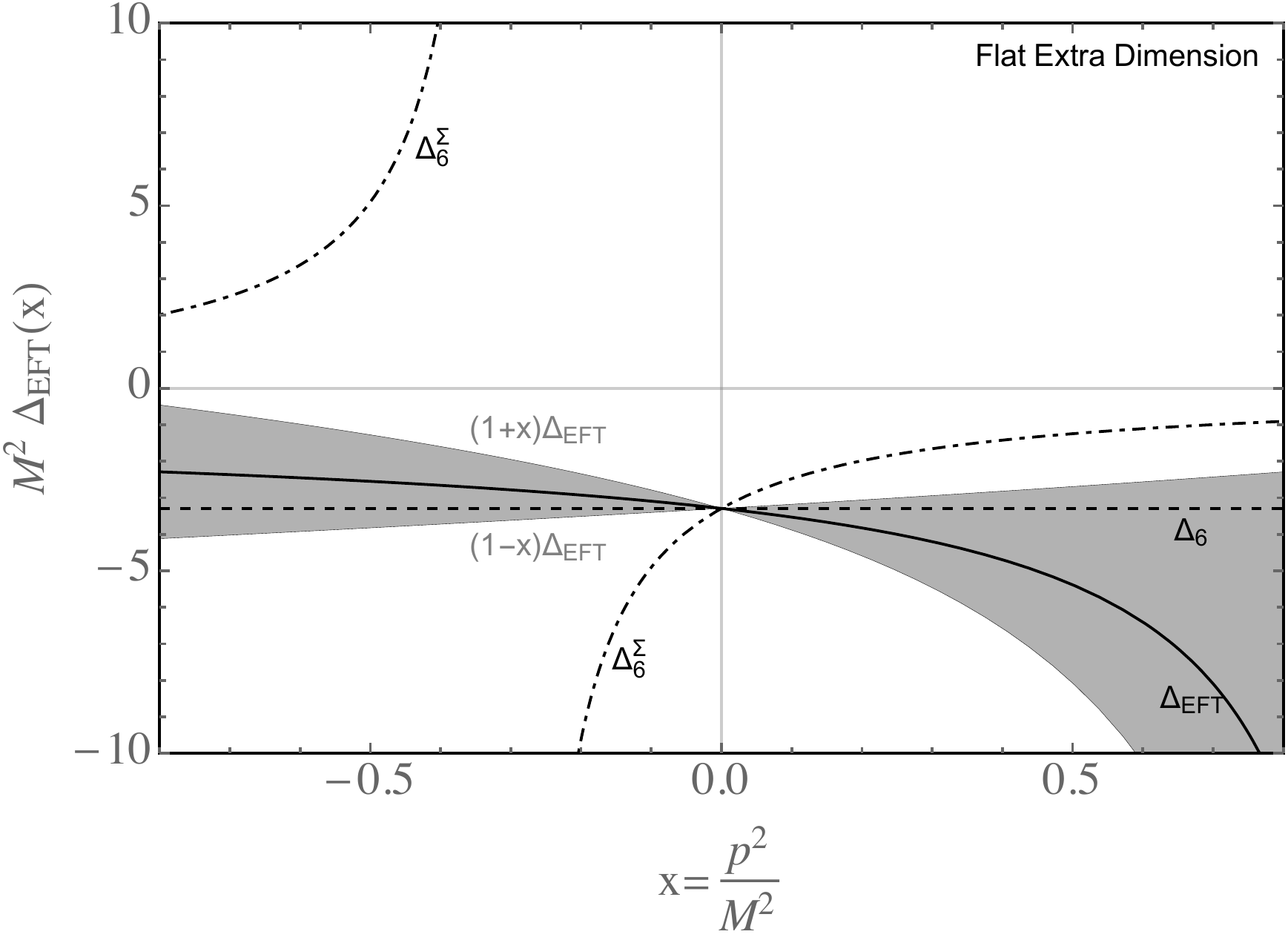}
\caption{The magnitude of EFT corrections to the propagator in the flat extra dimension Universal theory.  Here we plot the full EFT correction to the propagator $\Delta_{\text{EFT}}(x)$ as compared to the using the self-energy approach $\Delta^\Sigma_6(x)$, for which the derivative series diverges, and the approach advertised here, in which the propagator is expanded consistently at dimension-6 to find the correction $\Delta_6(x)$ wherein the derivative expansion must converge.  We also show an envelope around the full correction within which the dimension-6 approximation is expected to fall.}
\label{fig:validity}
\end{center}
\end{figure}

Now we concentrate on the gauge bosons, which describe the flat extra dimensional example of Universal theory given in \cite{Barbieri:2004qk}.  In this $\alpha \to 0$ limit one has
\be
\Delta (x) =  \frac{\pi}{M^2} \frac{\cot \left( \pi \sqrt{x}\right)}{\sqrt{x}} ~~.
\ee
In terms of the cutoff scale for the gauge boson propagator $M=1/R$ the Wilson coefficients of the expansion are
\be
c_1 = \frac{\pi^2}{3} ~~~,~~~ c_2 = \frac{\pi^2}{15} c_1 ~~~,~~~ c_3 = \frac{2 \pi^2}{21} c_2~~~,~~...~~,
\ee
\be
{\hat c}_1 = \frac{\pi^2}{3} ~~~,~~~ {\hat c}_2 = \frac{2\pi^2}{5} {\hat c}_1  ~~~,~~~ {\hat c}_3 = \frac{17 \pi^2}{42} {\hat c}_2~~~,~~...
\ee
Note that $c_1>1$ and thus, as expected from the general discussion in \Sec{sec:self}, the coefficients ${\hat c}_n$ of the self-energy expansion grow rapidly, while the coefficients $c_n$ satisfy `convergence'.  This provides a clear example of a situation in which, if working at dimension-6, one should expand the propagator to dimension-6 and not retain the dimension-6 term in the denominator of the propagator since this implicitly includes, at dimension-8, a term proportional to $c_1^2$ which is a factor $5$ larger than the true dimension-8 term of the full EFT.  This limitation of working with the self-energy in Universal theories is illustrated \Fig{fig:validity} where we have defined
\be
\Delta_{\text{EFT}}(x) = \Delta (x) - \frac{1}{M^2 x} ~~,~~ \Delta_6(x) = \frac{c_1}{M^2} ~~,~~ \Delta^\Sigma_6(x) = \frac{1}{M^2} \left( \frac{1}{x-{\hat c}_1 x^2} - \frac{1}{x} \right) ~~.
\ee

We see in \Fig{fig:validity} that the `self-energy' approach consistently fails to provide a good approximation to the full EFT result.  This is most notable at negative $x$, as found for $t$-channel exchange diagrams, in which an unphysical pole appears well below the true cutoff of the EFT and beyond this pole the `self-energy' approach even predicts an incorrect sign for the amplitude correction.

This illustrative example is only a toy model for a number of reasons.  Most notably is that the bulk does not respect custodial symmetry hence large violations of low energy precision electroweak constraints are possible.  Furthermore, all of the fine-tuning considerations relevant to flat extra-dimensional models will apply here, meaning that the Higgs is not necessarily naturally light.  Nonetheless, this simple examples demonstrates that a wide range of Wilson coefficients may be possible in examples of Universal theories, showing that it will be important to measure all electroweak oblique parameters to fully map the space of UV theories, and furthermore shows that for Universal theories if the propagator is not consistently expanded at the appropriate dimension the momentum expansion can break down prematurely, invalidating the use of the EFT.

\subsubsection*{An example for large {\boldmath $\hat{H}$}}
It is also straightforward to find examples where all of the `Higgs-only' operators, again including $\mathcal{O}_\Box$, arise at leading order, whereas the ones involving the gauge field strengths arise one loop higher in perturbation theory.  

A concrete example is a two-Higgs doublet model with all scalar sector couplings included, which may also be extended with an additional complex scalar singlet.  We may write this class of UV-completions as
\be
\mathcal{L} = \mathcal{L} _{\text{SM}} + |D_\mu \widetilde{H}|^2+ |\partial_\mu \widetilde{S} |^2 + \kappa \left( D_\mu \widetilde{H} D^\mu H + {\rm h.c.} \right) - V(H,\widetilde{H},\widetilde{S}) ~~,
\label{eq:lagg1}
\ee
where $\mathcal{L} _{\text{SM}}$ is the SM Lagrangian including kinetic and Yukawa terms for the SM-like Higgs doublet $H$. To avoid ghosts we take $|\kappa|\leqslant 1$, and the potential $V$ includes a mass term parameterised as
\be
V_{\text{mass}} = m^2 \widetilde{H}^\dagger \widetilde{H} + \left(   \beta m^2 \widetilde{H}^\dagger H + {\rm h.c.} \right) ~~,
\label{eq:lagg2}
\ee
as well as scalar interactions with a typical coupling strength $g_\ast$.  As expected from the coupling dimensions shown in \Tab{table:Mixed}, as one takes the limit $g_\ast \to 0$ the theory generates only $\mathcal{O}_\Box$ at leading order.  

At low energies, we can integrate out $\widetilde{H}$ by using its equations of motion, finding an effective theory described by 
\be
\mathcal{L}_{\text{EFT}} =  \mathcal{L} _{\text{SM}} + H^\dagger \, \frac{\left( \kappa \Box + \beta m^2 \right)^2}{\Box+m^2}\, H ~~.
\ee
After correcting for wave-function and mass rescaling, the tower of higher-dimension operators for a canonically normalised Higgs field is
\be
\mathcal{L}_{\text{EFT}} =  \frac{(\beta - \kappa)^2}{m^2(1+ \beta^2-2 \kappa \beta )} \, \, \Box H^\dagger \sum_{n=0}^{\infty} \left( \frac{-\Box}{m^2}\right)^n \Box H    \, .
\label{eq:WilsonOBox}
\ee
Thus we have presented an example of UV theory in which $\mathcal{O}_\Box$ emerges at low energy as a leading effect, giving
\be
{\hat H} = \frac{(\beta - \kappa)^2\, m_h^2}{(1+ \beta^2-2 \kappa \beta )\, m^2} ~.
\label{finH}
\ee
As expected, $\hat{H}$ turns out to be positive (for $|\kappa|\leqslant 1$). Of course, by turning on the coupling $g_\ast$, the other `Higgs-only' operators will be generated as well.

To compare to the mass scale of new physical states we can start from the theory described by eqs.~(\ref{eq:lagg1})--(\ref{eq:lagg2}) and diagonalise the $H$--$\widetilde{H}$ system.  After diagonalising, the heavy scalar has mass
\be
M^2 = \frac{1+ \beta^2-2 \kappa \beta}{1-\kappa^2} \, m^2 ~~,
\ee
where we have chosen the Higgs mass-squared such that a massless Higgs-like scalar remains.\footnote{This simplification is taken only to remove some parametric freedom, but one can easily include the non-zero Higgs mass.} 
Expressing the derivative expansion in terms of the physical cutoff mass $M$, we obtain from \Eq{eq:WilsonOBox} the Wilson coefficients
\be
{\hat c}_n = \frac{(\beta - \kappa)^2(1+\beta^2-2 \kappa \beta)^{n-1}}{(1-\kappa^2)^n}~.
\ee
Using the relation in \Eq{eq:hatcn} for $p^2\gg m_h^2$, we find
\be
c_n= \frac{(\beta - \kappa)^2}{1-\kappa^2} ~,~~~~{\hat H} = \frac{c_1\, m_h^2}{M^2} ~.
\ee
As expected from our general discussion, the coefficients $c_n$ satisfy positivity, convergence and are all equal, corresponding to tree-level single-particle exchange.

\subsection{EFT validity}
\label{sec:validity}

By construction, the range of EFT validity is up to energies of order $M$. As discussed in \Sec{sec:self}, the property of convergence is crucial to assess the correct interpretation of the extent of the EFT validity. However, through low-energy measurements we cannot determine the cutoff mass $M$ and the Wilson coefficient $c_\Box$ separately, but only in the combination $c_\Box /M^2 \equiv \Hat H / m_h^2$ that appears in the definition of $\hat{H}$. As the oblique parameter $\hat{H}$ leads to energy-growing effects, one is interested to know what is the maximum energy for which the EFT prediction can be trusted when compared with an experimental measurement. For a given value of $\hat{H}$, the maximum value of the EFT cutoff corresponds to the maximum possible value of the coefficient $c_\Box$. Therefore, the question of the range of the EFT validity translates into a question about the maximum value of $c_\Box$. 

A na\"ive upper bound on $c_\Box$ can be obtained by requiring that the coupling in \Eq{eq:cappa} must satisfy a generic perturbative bound $\kappa < 4\pi$. This motivates the limit 
\be
c_\Box \lesssim (4 \pi)^2 ~,
\label{eq:limit}
\ee
which corresponds to the request that the maximum energy for which the EFT prediction can be trusted is 
\be
E_{\rm max} \lesssim  \frac{4\pi }{\sqrt{\hat{H}}}\, m_h ~.
\label{eq:limitE}
\ee
We will refer to eqs.~(\ref{eq:limit})--(\ref{eq:limitE}) as the \emph{na\"ive perturbativity} constraint, since the UV-completion which violates these simple bounds is likely to be non-perturbative. 

In general, the na\"ive perturbativity constraint is over-optimistic and, possibly, unrealistic. This is because the corresponding value of $c_\Box$ likely violates perturbative unitarity, as applied to some scattering process, both within the EFT itself or in the underlying UV-completion.  One particularly constraining process is $t \bar{t} \to t \bar{t}$ scattering mediated by an off-shell Higgs.  In this case, leading order perturbative unitarity is typically not violated within the regime of validity of the EFT ($p^2 < M^2$) whenever
\be
|c_\Box| \lesssim 4 \pi ~,
\label{eq:limit2}
\ee
where the precise coefficient depends on the specific process under consideration. The corresponding limit on the maximum energy for which the EFT can be trusted is
\be
E_{\rm max} \lesssim  \sqrt{\frac{4\pi}{\hat{H}}} \, m_h ~.
\label{eq:limitE2}
\ee
We will refer to eqs.~(\ref{eq:limit2})--(\ref{eq:limitE2})  as the \emph{perturbative unitarity} constraint.  

Both na\"ive perturbativity and perturbative unitarity provide useful, although qualitative, constraints to guide our phenomenological study of the $\hat{H}$-parameter.

\section{Probing {\boldmath $\hat{H}$} at colliders}
\label{sec:pheno} 
In this section we will discuss how high-energy colliders can search for the Higgs oblique parameter $\hat{H}$.

 \subsection{On-shell probes}
 \label{sec:onshell}
 
 As shown in \Sec{sec:physeff}, the oblique parameter $\hat{H}$ affects the on-shell Higgs couplings only with a universal modification of the interaction to fermions (usually parametrised by the coefficient $\kappa_f$)
 \be
 \kappa_f = 1-\hat{H} ~.
 \ee
 We recall that the positivity condition discussed in \Sec{sec:KL} requires $\hat{H} \geqslant  0$, so $\kappa_f$ is always reduced with respect to the SM value.
The latest combined fit of the ATLAS collaboration on fermionic Higgs couplings, involving both Higgs production and decay processes and using up to $80$~fb$^{-1}$ of $13$~TeV data~\cite{ATLAS:2018doi}, gives
\be
\hat{H} < 0.16 ~~~\textnormal{at 95\% CL ~~~ (LHC~today)},
\ee
where the bound is obtained by assuming that  $\kappa_f$ is the only new-physics effect in Higgs physics.
Recent estimates of the projections of Higgs coupling measurements at the HL-LHC with 3~ab$^{-1}$~\cite{ATL-PHYS-PUB-2018-054} translate into a future bound
\be
\hat{H} < 0.04 ~~~\textnormal{at 95\% CL ~~~ (HL-LHC projection)}.
\label{eq:hllhc}
\ee

\subsection{Off-shell probes}
\label{sec:off-shell-probes}

Off-shell Higgs exchange can affect a physical process with contributions that, at the amplitude level,
scale as $ \hat{H} \, p^2/m_h^2$.  Thus, even if the measurement of such a process at high energies is not as precise as a low-energy measurement, it may still be competitive with high precision low-energy constraints, such as those from on-shell observables. Moreover, while the reach of on-shell probes given in \Eq{eq:hllhc} offers a useful benchmark, we stress that off-shell probes should be carried out independently. Indeed, as shown in \Sec{sec:physeff}, contributions from other operators generally present in a Universal EFT can affect, and even cancel out, modifications of SM Higgs couplings. On the contrary, the search for off-shell effects is a unique and clean test of the Higgs oblique parameter $\hat{H}$.

As shown in \Sec{sec:physeff}, the study of the process $pp \to h^\ast \to VV$ is futile for testing $\hat{H}$, since its energy-growing effects exactly vanish in the corresponding amplitude. The next obvious place to look for energy-growing contributions in proton colliders is $t \bar{t}$ production mediated by an off-shell Higgs. However, while the signal comes from a loop-induced process, the $t \bar{t}$ SM background is a tree-level QCD process. Thus, this channel gives an inefficient probe of $\hat{H}$.

Moreover, the $\hat{H}$ contribution to $t \bar{t}$ production comes from various one-loop Feynman diagrams, some of which contain a modified Higgs propagator inside the loop. This can potentially lead to a logarithmic sensitivity on the cut-off which obscures the data interpretation and introduces a model dependence.

The next process to consider is Higgs pair production.  In this case, the modified Higgs propagator does not run inside the loop and there is no model-dependent cutoff sensitivity.  However, the cross section falls rapidly due to the top-loop form factor, and this counteracts the energy-growing behaviour from $\hat{H}$.  For instance, the total di-Higgs cross section at the 14 TeV LHC, with the cut $m_{hh} < 1.5$ TeV, is modified by $\hat{H}=0.04$ at the $23\%$ level.  Given the limited sensitivity to Higgs pair production at the HL-LHC, this channel is unlikely to be competitive with on-shell constraints on the $\hat{H}$-parameter at the LHC.

It transpires that the most promising channel for off-shell probes of $\hat{H}$ is a more exotic process: four-top production. 

\subsubsection*{Four-top production}

\begin{figure}[tbp]
\begin{center}
\includegraphics[width=0.55\textwidth]{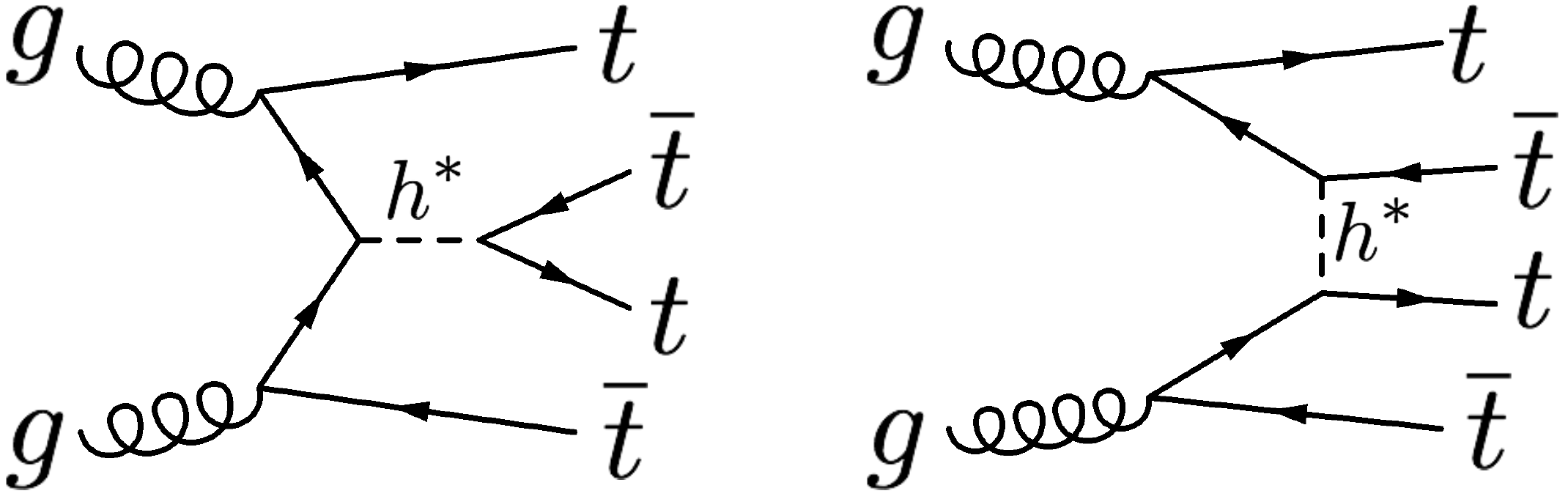}
\caption{A sample of Feynman diagrams with an off-shell Higgs contribution to four-top production at the LHC ($p p \to t \bar t t \bar t$).}
\label{fig:diagrams4t}
\end{center}
\end{figure}

\begin{figure}[tbp]
\begin{center}
\includegraphics[width=0.75\textwidth]{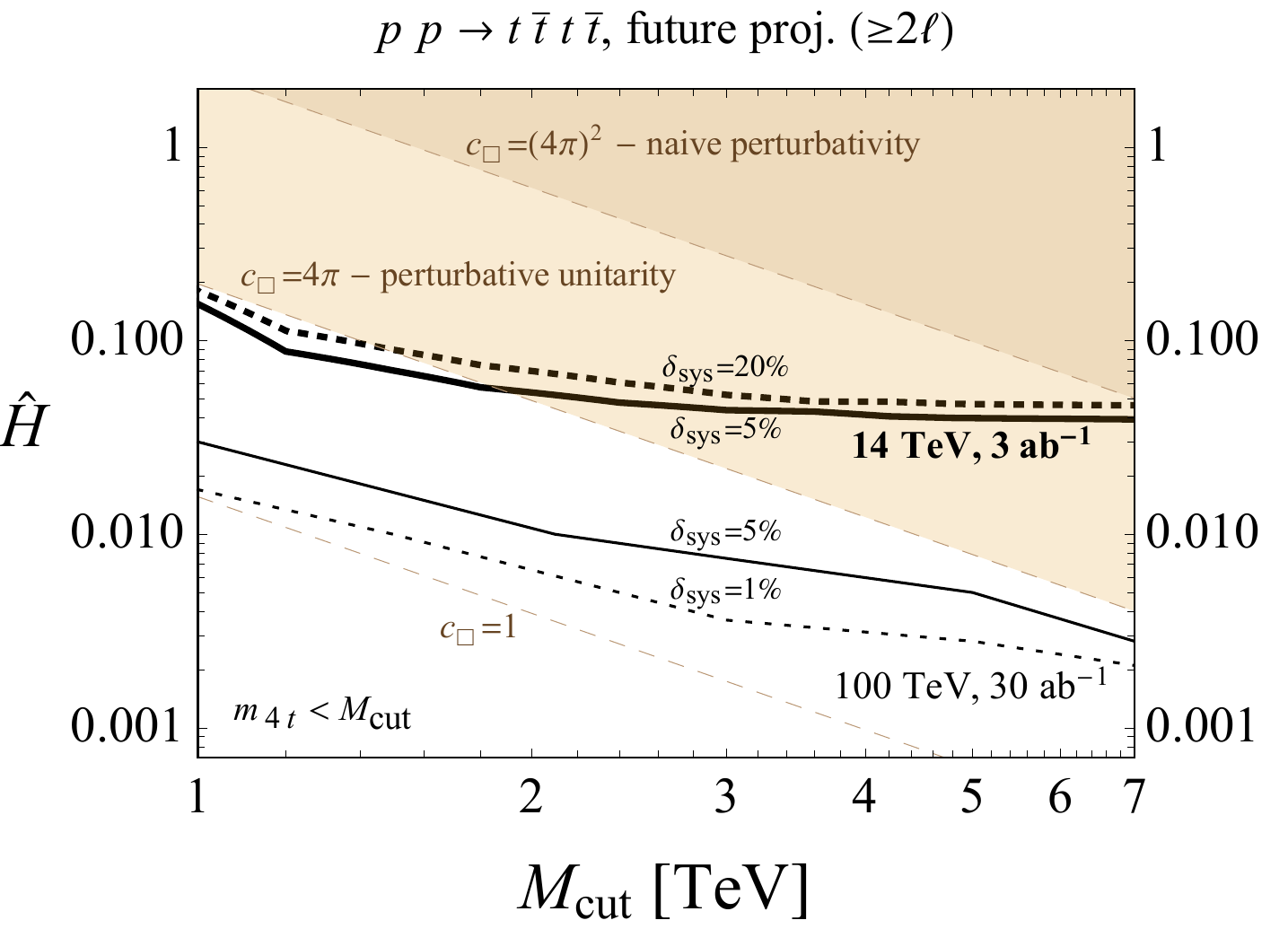}
\caption{The 3~ab$^{-1}$ HL-LHC and  30~ab$^{-1}$ FCC-hh sensitivity projections for the $\hat H$ parameter in four-top production ($p p \to t \bar t t \bar t$).  The solid and dashed black curves show the expected sensitivity at 95\% CL as a function of the kinematic variable $M_{\textrm{cut}}$ for a different systematic uncertainty $\delta_{sys}$. Superimposed to this plot are three dashed brown lines showing the corresponding values of $c_\Box$ assuming $M = M_{\textrm{cut}}$. The regions above the lines $c_\Box =4\pi$ and $c_\Box =(4\pi)^2$ are incompatible with the criteria of {\it perturbative unitarity} and {\it na\"ive perturbativity}, respectively.}
\label{fig:results4t}
\end{center}
\end{figure}

Here we consider the role of the process $p p \to t \bar{t} t \bar{t}$ as a probe of the Higgs boson off-shell, see \Fig{fig:diagrams4t}. Four-top production at the LHC is a rare process in the SM with cross section of $15.8 \pm 3.1$\,fb at 14~TeV collider energy (NLO QCD + EWK)~\cite{Bevilacqua:2012em,Frederix:2017wme}. The dynamical scale choice $\mu_R = \mu_F = H_T /4$ is particularly effective in stabilising the distribution corrections from LO to NLO~\cite{Bevilacqua:2012em} and will be employed in the analysis below. Here, $H_T$ is defined as the total transverse energy of the four-top system, $H_T = \sum_{i=1}^{4} \sqrt{m_t^2 + p_T^2(t_i)}$.

Due to statistics, systematics and background, the four-top final state is challenging to observe~\cite{Alvarez:2016nrz}. Nonetheless, significant progress by the experimental collaborations has been made recently.  Both ATLAS and CMS analysed about 36~fb$^{-1}$ of 13~TeV data each~\cite{Aaboud:2018jsj,Sirunyan:2017roi}, with constraints approaching the SM rate. Interestingly, ATLAS reported comparable sensitivities in the combination of single lepton plus opposite-sign dilepton searches when compared to the combination of same-sign dilepton plus three lepton searches. The first class of searches selects more signal events but suffers from larger systematic uncertainties. In fact, these are already now becoming a limiting factor. Therefore, to derive future projections we will focus on the second class of searches, which feature rarer but cleaner signatures.

ATLAS and CMS have also studied projections for four-top production at the HL-LHC~\cite{ATL-PHYS-PUB-2018-047,CMS-PAS-FTR-18-031} (see also~\cite{Azzi:2019yne}). Both reported the expected statistical uncertainty of $9\%$ on the SM signal strength modifier ($\mu = \sigma / \sigma_{\textrm{SM}}$). However, ATLAS quotes an expected sensitivity including systematics of $16\%$, while the CMS estimate ranges from $18\%$ to $28\%$. The major source of systematic uncertainty comes from the theoretical uncertainty on signal and background normalisation. Hence, it is reasonable to expect that improved theoretical calculations of $p p \to t \bar t t \bar t$, $t \bar t V$ and $t \bar t H$ will considerably reduce this uncertainty. Nevertheless, to be conservative, here we show results for two benchmark scenarios, $\delta_{\textrm{sys}} = 5 \%$ and $\delta_{\textrm{sys}} = 20 \%$.

The same-sign dilepton and trilepton projection analysis by ATLAS~\cite{ATL-PHYS-PUB-2018-047} exploits three particularly clean categories with at least 6 jets, out of which 3 or 4 are b-jets, yielding $S/\sqrt{B} \sim 10$ and $S/B$ in the range of 2.3 to 5.5. The total expected number of events  in these categories at 14~TeV and 3~ab$^{-1}$ is about $120$.

We use \MadGraph\cite{Alwall:2014hca} to perform leading-order parton-level studies of $p p \to t \bar t t \bar t$ including the Higgs oblique parameter $\hat{H}$. We adjusted the SM \UFO model files to incorporate the Higgs boson propagator modification -- according to \Eq{eq:propagator} -- and the modification of the top Yukawa interaction -- keeping only the $\hat{H}$ correction in \Eq{eq:yukyuk}.
To cross check the results in a different field basis we also implemented an equivalent modified top Yukawa and four-top operator in the \FeynRules~\cite{Alloul:2013bka}, exported to \UFO, and confirmed agreement between the two procedures. We find that at 14 TeV the fractional modification to the inclusive $t \bar{t} t \bar{t}$ production cross section is 
\be
\delta \sigma_{t \bar{t} t \bar{t}} \equiv \frac{\sigma_{\hat{H}}-\sigma_{\text{SM}}}{\sigma_{\text{SM}}} 
\approx 0.03 \, \left( \frac{\hat{H}}{0.04} \right) + 0.15 \, \left( \frac{\hat{H}}{0.04} \right)^2 ~,
\ee
showing competitive sensitivity to the on-shell probes already at this level. The interference effects between SM and $\hat H$-induced diagrams are sub-leading given the expected experimental reach. 

We perform kinematical cuts in two variables, $H_T$ and $m_{4 t}$, both of which can be reasonably well-approximated in a realistic analysis setup. Here, $m_{4 t}$ is the total invariant mass of the four-top system, while $H_T$ is the total transverse energy defined before. We have checked explicitly that the simulated events satisfy $| p_h^2 | < m_{4 t}^2$, where $|p_h^2|$ is the maximal momentum flow in the Higgs propagator for all Feynman diagrams.

Shown in \Fig{fig:results4t} is the expected sensitivity (at $95\%$ CL) on $\hat H $ for a given upper limit on $m_{4 t} \leq M_{\textrm{cut}}$, after optimising the $H_T$ cut. The number of events in the final selection bin is described with Poisson distribution. The black solid (dashed) line corresponds to the overall systematic uncertainty of 5\% (20\%). We repeat this exercise with the exact same procedure for $100$~TeV proton-proton collider and $30$~ab$^{-1}$ of luminosity, assuming the systematic uncertainties of 5\% and 1\%, respectively.  Based on extrapolations of higher order perturbative calculations a $5\%$ systematic error seems realistic, whereas $1\%$ may be optimistic, depending on future progress.

To assess the reliability of the EFT prediction in the plane of \Fig{fig:results4t} we recall the discussion in \Sec{sec:validity}. Since the energy flowing in the Higgs propagator never exceeds $M_{\textrm{cut}}$, we can interpret $M_{\textrm{cut}}$ as the minimum possible value of the EFT cutoff and therefore $c_\Box \geqslant \hat{H} M_{\textrm{cut}}^2 /m_h^2$. We then plot in \Fig{fig:results4t} the corresponding values of $c_\Box$, identifying the regions in conflict with the criteria of {\it perturbative unitarity} ($c_\Box \gtrsim 4\pi$) and {\it na\"ive perturbativity} ($c_\Box \gtrsim (4\pi)^2$).

 To summarise \Fig{fig:results4t}, future HL-LHC four-top searches will provide a competitive probe of $\hat{H}$ in the off-shell Higgs regime, giving meaningful constraints on a wide class of theories featuring moderate to strong coupling constants. The FCC-hh collider has a potential to probe weakly coupled theories and, at large cutoff, potentially supersede the FCC-ee precision constraint on an $\hat{H}$-only scenario, which would be at the level of $|\hat{H}|\lesssim 0.5 \%$ \cite{Mangano:2018mur}.

While this simple analysis already illustrates the importance of the four-top production in the context of Higgs physics, it is far from unlocking the full potential of this process. We envisage a number of possible improvements. For example, $t \bar{t} t \bar{t}$ angular distributions could help disentangle signal from the background. In this context, we identify a suitable parton-level variable, $\Delta =  \eta_{t_1} + \eta_{t_2} -\eta_{\bar{t}_1} - \eta_{\bar{t}_2}$ which could be employed to further enhance the sensitivity. However, a realistic collider analysis is beyond the scope of this paper and the simulation of decays, showering, hadronisation and detector effects, possibly employing advanced machine learning techniques for optimised results, is left for future work.

\section{Conclusions}
The future of Higgs physics will have a course charted by precision calculations and a destination mapped by a new frontier of experimental measurements.  The resulting landscape will be translated into fundamental questions:  What is the nature of the Higgs boson? How does the Higgs boson interact with other particles and with itself?  In this work we have advertised and studied an orthogonal, yet important, question for this programme:  How does the Higgs boson propagate?  Framed within a general EFT context the answer to this question is unphysical and basis-dependent.  However there is a broad class of microscopic theories (called Universal theories) which single out a specific EFT basis in which this question not only becomes well-defined, but also plays a key role in mapping out the boundaries of the UV.  Leading order modifications of the Higgs propagator are captured by the $\hat{H}$-parameter, which is the coefficient of the operator $\mathcal{O}_\Box = |\Box H|^2$ in the Universal basis. The $\hat{H}$-parameter provides a Higgs-boson analogue to the oblique electroweak parameter programme and, since it measures the high-momentum corrections to the propagator, thus is the hallmark of off-shell Higgs physics.

In \Sec{sec:KL} we set course by studying the general properties of propagators in QFT. Starting from the non-perturbative \KL~representation, we derive some consistency conditions that must be satisfied by the Wilson coefficients of the EFT expansion.
In particular, we discuss a {\it positivity} condition for the coefficients of the two-point function and a so-called {\it convergence} condition, governing the relation between successive coefficients. Convergence can be used to place upper bounds on the scale of new states if successive Wilson coefficients are measured.  With regard to the Higgs boson, the \KL~representation can be used to constrain the sign of the $\hat{H}$-parameter in a very broad range of UV-completions.

Even within the limited territory of Universal EFTs, in \Sec{sec:theory} it was shown that the physical effects of the $\hat{H}$-parameter cannot be unambiguously constrained by on-shell Higgs coupling measurements alone. Off-shell Higgs physics becomes the natural arena to test the oblique $\hat{H}$-parameter.
This promotes precision measurements involving an off-shell Higgs boson to a key exploratory role within the precision Higgs era. The off-shell processes provide information that cannot be accessed simply with on-shell measurements and is crucial to break degeneracies between Wilson coefficients in order to fully explore the space of Universal EFTs.
To illustrate the possibilities to which such measurements are sensitive, a small sample of UV possibilities were discussed in \Sec{sec:UV}.

Finally, after exploring a variety of different off-shell processes and showing that energy-growing effects in $gg\to h^\ast\to VV$ cancel exactly, in \Sec{sec:pheno} four-top production was demonstrated to be a promising probe of the $\hat{H}$-parameter, competing quantitatively with on-shell coupling measurements for moderately and strongly-coupled microscopic models.  In conclusion, future HL-LHC studies of four-top production would provide important complementary information on Higgs-sector modifications arising in a wide range of microscopic theories, forming a crucial component in the wider effort to determine the microscopic nature of electroweak symmetry breaking.

%%%%%%%%%%%%%%%%%%%%%%%%%%%%%%%%%%%%%%%%%%%%%%%%%%%%%%%%%%%%%%%%%%%%%%

\bigskip\bigskip
\acknowledgments
We are grateful to Brando Bellazzini, Joan Elias-Mir\'{o}, Mart\'{i}n Gonzalez-Alonso,  Brian Henning, Michelangelo Mangano, Marco Nardecchia, Riccardo Rattazzi, Francesco Riva, Yotam Soreq, Jesse Thaler, and Andrea Wulzer for stimulating conversations.  CE is supported by the UK Science and Technology Facilities Council (STFC) under grant ST/P000746/1.

\appendix
\section{Bernoulli, Veneziano, and {\boldmath $\pi$}}
\label{app:fun}
Suppose we have a general form of a propagator $\Delta(s)$ or forward scattering amplitude $\mathcal{M}$, which may be described by $l$-subtracted dispersion relations of the form
\be
\Delta(s) =  \int^\infty_{0} dq^2 \frac{\rho_\mathcal{O}(q^2)}{s-q^2+i \epsilon} + \text{Poly}(s) ~~,
\ee
and
\be
\mathcal{M} (s) =  \int^\infty_{0} dq^2  \left( \frac{F(q^2)}{s+q^2+i \epsilon}- \frac{F(q^2)}{s-q^2-i \epsilon}\right)  +\text{Poly}(s) ~~,
\ee
where any poles or branch cuts begin at some fixed scale $M$, such that $\rho_\mathcal{O}(q^2<M^2)=0$ and $F(q^2<M^2)=0$. Here $\text{Poly}(s)$ is a polynomial function of $s$ up to order $l-1$, with coefficients chosen to render the final result finite and consistent with observations.  Let us define
\be
a_{n>l} = \frac{1}{n!} \frac{d^n \Delta (s)}{d s^n} \bigg|_{s=0} ~~~~,~~~~ b_{n>l/2} = \frac{1}{2n!} \frac{d^{2n} \mathcal{M} (s)}{d s^{2n}} \bigg|_{s=0} ~~,
\ee
where in both instances we implicitly assume enough derivatives such that the subtractions are no longer relevant.  Consider the ratios
\be
T_n = M^2 \frac{a_{n+1}}{a_{n}} ~~~~,~~~~ R_n = M^4 \frac{b_{n+1}}{b_{n}} ~~.
\ee
From the dispersion relation we have the convergence condition
\be
T_{n} \leqslant 1 ~~~~,~~~~  R_{n} \leqslant 1 ~~.
\ee
Furthermore, for $\rho_\mathcal{O}(q^2)$ and $F(q^2)$ which grow sufficiently slowly, as may be determined from, for example, the optical theorem and the Froissart bound, we observe the limiting behaviour
\be
T_{n\to \infty} \to 1 ~~~~,~~~~  R_{n\to \infty} \to 1 ~~.
\ee
This has an important consequence, which is that as we take the limit $n\to \infty$ then, for any dispersion relation, including those involving loops or strongly coupled sectors, the Wilson coefficients must asymptotically approach the value for tree-level exchange.

As an amusing application of this observation, consider the tree-level scattering amplitude for gauge boson scattering in string theory at lowest order in the $g_s$ expansion \cite{Polchinski:1998rr,Adams:2006sv}
\be
\mathcal{A} \propto g_s K(\epsilon_i,p_i) \left[ \frac{\Gamma(-\alpha' s)\Gamma(-\alpha' u)}{\Gamma(1-\alpha' s-\alpha' u)}  + (s\to t) + (u\to t) \right] ~~,
\ee
where the string scale is $M_S^2 = 1/\alpha'$  and the functional form is proportional to the Veneziano amplitude.  The forward limit is\footnote{To confirm that this amplitude may be written with the desired dispersion relation, using the identity 
\be
\pi \cot(\pi x) = \lim_{N\to \infty} \sum_{n=-N}^N \frac{1}{x+n} ~~,
\ee
we find that this forward amplitude is described by a once-subtracted dispersion relation with
\be
F(q^2) \propto \sum_{n=0}^\infty (2 n+1) \delta\left(q^2 -  \frac{2n+1}{\alpha'} \right) ~~.
\ee
}
\be
\mathcal{M} (s) \propto s \tan \left(\alpha' \frac{\pi s}{2} \right) ~~.
\ee
Expanding this forward amplitude we have
\be
R_{n} = -\frac{\pi^2}{(2 n+2) (2 n +1)} \frac{2^{2n+2}-1}{2^{2 n}-1} \frac{\mathcal{B}_{2n+2}}{\mathcal{B}_{2n}} ~~.
\ee
Hence, from convergence, we find upper bounds on ratios of Bernoulli numbers
\be
 -\frac{\mathcal{B}_{2n+2}}{\mathcal{B}_{2n}} \leqslant \frac{(2 n+2) (2 n +1)}{\pi^2} \frac{2^{2 n}-1}{2^{2n+2}-1}  ~~,
 \label{eq:bern1}
\ee
and from the convergence limit we can connect the rational Bernoulli numbers to the irrational number $\pi$ as 
\be
\lim_{n\to \infty} \,  (2 n+2) (2 n +1) \frac{2^{2 n}-1}{2^{2n+2}-1} \frac{\mathcal{B}_{2n}}{\mathcal{B}_{2n+2}}  = -\pi^2 ~~.
\label{eq:bern2}
\ee
The identity in \Eq{eq:bern2} is a well-known result in number theory and the inequality (\ref{eq:bern1}) has been recently obtained in ref.~\cite{Qi}. It is curious that one can turn around the argument and find these two results on Bernoulli numbers starting from the convergence criterion applied to the EFT expansion of the Veneziano amplitude.

{Note that we have only worked at leading order in $g_s$, however higher order corrections would likely also contribute.\footnote{We thank Brando Bellazzini for discussions on this aspect.}  At $\mathcal{O}(g_s^2)$ one also has a contribution from closed string exchange, which includes a t-channel singularity from the massless graviton.  Since we are concerned with higher orders in $s$ in the forward limit this singularity does not affect the discussion above some low power in $s_n$.}

\bibliographystyle{JHEP}
\bibliography{references}

\providecommand{\href}[2]{#2}\begingroup\raggedright\begin{thebibliography}{10}

\bibitem{Peskin:1990zt}
M.~E. Peskin and T.~Takeuchi, \emph{{A New constraint on a strongly interacting
  Higgs sector}}, \href{https://doi.org/10.1103/PhysRevLett.65.964}{\emph{Phys.
  Rev. Lett.} {\bfseries 65} (1990) 964--967}.

\bibitem{Golden:1990ig}
M.~Golden and L.~Randall, \emph{{Radiative Corrections to Electroweak
  Parameters in Technicolor Theories}},
  \href{https://doi.org/10.1016/0550-3213(91)90614-4}{\emph{Nucl. Phys.}
  {\bfseries B361} (1991) 3--23}.

\bibitem{Holdom:1990tc}
B.~Holdom and J.~Terning, \emph{{Large corrections to electroweak parameters in
  technicolor theories}},
  \href{https://doi.org/10.1016/0370-2693(90)91054-F}{\emph{Phys. Lett.}
  {\bfseries B247} (1990) 88--92}.

\bibitem{Altarelli:1990zd}
G.~Altarelli and R.~Barbieri, \emph{{Vacuum polarization effects of new physics
  on electroweak processes}},
  \href{https://doi.org/10.1016/0370-2693(91)91378-9}{\emph{Phys. Lett.}
  {\bfseries B253} (1991) 161--167}.

\bibitem{Grinstein:1991cd}
B.~Grinstein and M.~B. Wise, \emph{{Operator analysis for precision electroweak
  physics}}, \href{https://doi.org/10.1016/0370-2693(91)90061-T}{\emph{Phys.
  Lett.} {\bfseries B265} (1991) 326--334}.

\bibitem{Peskin:1991sw}
M.~E. Peskin and T.~Takeuchi, \emph{{Estimation of oblique electroweak
  corrections}}, \href{https://doi.org/10.1103/PhysRevD.46.381}{\emph{Phys.
  Rev.} {\bfseries D46} (1992) 381--409}.

\bibitem{Altarelli:1991fk}
G.~Altarelli, R.~Barbieri and S.~Jadach, \emph{{Toward a model independent
  analysis of electroweak data}},
  \href{https://doi.org/10.1016/0550-3213(92)90376-M}{\emph{Nucl. Phys.}
  {\bfseries B369} (1992) 3--32}.

\bibitem{Burgess:1993mg}
C.~P. Burgess, S.~Godfrey, H.~Konig, D.~London and I.~Maksymyk, \emph{{A Global
  fit to extended oblique parameters}},
  \href{https://doi.org/10.1016/0370-2693(94)91322-6}{\emph{Phys. Lett.}
  {\bfseries B326} (1994) 276--281},
  [\href{https://arxiv.org/abs/hep-ph/9307337}{{\ttfamily hep-ph/9307337}}].

\bibitem{Maksymyk:1993zm}
I.~Maksymyk, C.~P. Burgess and D.~London, \emph{{Beyond S, T and U}},
  \href{https://doi.org/10.1103/PhysRevD.50.529}{\emph{Phys. Rev.} {\bfseries
  D50} (1994) 529--535},
  [\href{https://arxiv.org/abs/hep-ph/9306267}{{\ttfamily hep-ph/9306267}}].

\bibitem{Barbieri:2004qk}
R.~Barbieri, A.~Pomarol, R.~Rattazzi and A.~Strumia, \emph{{Electroweak
  symmetry breaking after LEP-1 and LEP-2}},
  \href{https://doi.org/10.1016/j.nuclphysb.2004.10.014}{\emph{Nucl. Phys.}
  {\bfseries B703} (2004) 127--146},
  [\href{https://arxiv.org/abs/hep-ph/0405040}{{\ttfamily hep-ph/0405040}}].

\bibitem{Farina:2016rws}
M.~Farina, G.~Panico, D.~Pappadopulo, J.~T. Ruderman, R.~Torre and A.~Wulzer,
  \emph{{Energy helps accuracy: electroweak precision tests at hadron
  colliders}},
  \href{https://doi.org/10.1016/j.physletb.2017.06.043}{\emph{Phys. Lett.}
  {\bfseries B772} (2017) 210--215},
  [\href{https://arxiv.org/abs/1609.08157}{{\ttfamily 1609.08157}}].

\bibitem{Franceschini:2017xkh}
R.~Franceschini, G.~Panico, A.~Pomarol, F.~Riva and A.~Wulzer,
  \emph{{Electroweak Precision Tests in High-Energy Diboson Processes}},
  \href{https://doi.org/10.1007/JHEP02(2018)111}{\emph{JHEP} {\bfseries 02}
  (2018) 111}, [\href{https://arxiv.org/abs/1712.01310}{{\ttfamily
  1712.01310}}].

\bibitem{Banerjee:2018bio}
S.~Banerjee, C.~Englert, R.~S. Gupta and M.~Spannowsky, \emph{{Probing
  Electroweak Precision Physics via boosted Higgs-strahlung at the LHC}},
  \href{https://doi.org/10.1103/PhysRevD.98.095012}{\emph{Phys. Rev.}
  {\bfseries D98} (2018) 095012},
  [\href{https://arxiv.org/abs/1807.01796}{{\ttfamily 1807.01796}}].

\bibitem{Kallen:1952zz}
G.~Kallen, \emph{{On the definition of the Renormalization Constants in Quantum
  Electrodynamics}},
  \href{https://doi.org/10.1007/978-3-319-00627-7_90}{\emph{Helv. Phys. Acta}
  {\bfseries 25} (1952) 417}.

\bibitem{Lehmann:1954xi}
H.~Lehmann, \emph{{On the Properties of propagation functions and
  renormalization contants of quantized fields}},
  \href{https://doi.org/10.1007/BF02783624}{\emph{Nuovo Cim.} {\bfseries 11}
  (1954) 342--357}.

\bibitem{Adams:2006sv}
A.~Adams, N.~Arkani-Hamed, S.~Dubovsky, A.~Nicolis and R.~Rattazzi,
  \emph{{Causality, analyticity and an IR obstruction to UV completion}},
  \href{https://doi.org/10.1088/1126-6708/2006/10/014}{\emph{JHEP} {\bfseries
  10} (2006) 014}, [\href{https://arxiv.org/abs/hep-th/0602178}{{\ttfamily
  hep-th/0602178}}].

\bibitem{Cacciapaglia:2006pk}
G.~Cacciapaglia, C.~Csaki, G.~Marandella and A.~Strumia, \emph{{The Minimal Set
  of Electroweak Precision Parameters}},
  \href{https://doi.org/10.1103/PhysRevD.74.033011}{\emph{Phys. Rev.}
  {\bfseries D74} (2006) 033011},
  [\href{https://arxiv.org/abs/hep-ph/0604111}{{\ttfamily hep-ph/0604111}}].

\bibitem{Froissart:1961ux}
M.~Froissart, \emph{{Asymptotic behavior and subtractions in the Mandelstam
  representation}}, \href{https://doi.org/10.1103/PhysRev.123.1053}{\emph{Phys.
  Rev.} {\bfseries 123} (1961) 1053--1057}.

\bibitem{Bellazzini:2014waa}
B.~Bellazzini, L.~Martucci and R.~Torre, \emph{{Symmetries, Sum Rules and
  Constraints on Effective Field Theories}},
  \href{https://doi.org/10.1007/JHEP09(2014)100}{\emph{JHEP} {\bfseries 09}
  (2014) 100}, [\href{https://arxiv.org/abs/1405.2960}{{\ttfamily 1405.2960}}].

\bibitem{Low:2009di}
I.~Low, R.~Rattazzi and A.~Vichi, \emph{{Theoretical Constraints on the Higgs
  Effective Couplings}},
  \href{https://doi.org/10.1007/JHEP04(2010)126}{\emph{JHEP} {\bfseries 04}
  (2010) 126}, [\href{https://arxiv.org/abs/0907.5413}{{\ttfamily 0907.5413}}].

\bibitem{Falkowski:2012vh}
A.~Falkowski, S.~Rychkov and A.~Urbano, \emph{{What if the Higgs couplings to W
  and Z bosons are larger than in the Standard Model?}},
  \href{https://doi.org/10.1007/JHEP04(2012)073}{\emph{JHEP} {\bfseries 04}
  (2012) 073}, [\href{https://arxiv.org/abs/1202.1532}{{\ttfamily 1202.1532}}].

\bibitem{Urbano:2013aoa}
A.~Urbano, \emph{{Remarks on analyticity and unitarity in the presence of a
  Strongly Interacting Light Higgs}},
  \href{https://doi.org/10.1007/JHEP06(2014)060}{\emph{JHEP} {\bfseries 06}
  (2014) 060}, [\href{https://arxiv.org/abs/1310.5733}{{\ttfamily 1310.5733}}].

\bibitem{Bellazzini:2016xrt}
B.~Bellazzini, \emph{{Softness and amplitudes’ positivity for spinning
  particles}}, \href{https://doi.org/10.1007/JHEP02(2017)034}{\emph{JHEP}
  {\bfseries 02} (2017) 034},
  [\href{https://arxiv.org/abs/1605.06111}{{\ttfamily 1605.06111}}].

\bibitem{Tanabashi:2018oca}
{\scshape Particle Data Group} collaboration, M.~Tanabashi et~al.,
  \emph{{Review of Particle Physics}},
  \href{https://doi.org/10.1103/PhysRevD.98.030001}{\emph{Phys. Rev.}
  {\bfseries D98} (2018) 030001}.

\bibitem{Wells:2015uba}
J.~D. Wells and Z.~Zhang, \emph{{Effective theories of universal theories}},
  \href{https://doi.org/10.1007/JHEP01(2016)123}{\emph{JHEP} {\bfseries 01}
  (2016) 123}, [\href{https://arxiv.org/abs/1510.08462}{{\ttfamily
  1510.08462}}].

\bibitem{Giudice:2007fh}
G.~F. Giudice, C.~Grojean, A.~Pomarol and R.~Rattazzi, \emph{{The
  Strongly-Interacting Light Higgs}},
  \href{https://doi.org/10.1088/1126-6708/2007/06/045}{\emph{JHEP} {\bfseries
  06} (2007) 045}, [\href{https://arxiv.org/abs/hep-ph/0703164}{{\ttfamily
  hep-ph/0703164}}].

\bibitem{Grzadkowski:2010es}
B.~Grzadkowski, M.~Iskrzynski, M.~Misiak and J.~Rosiek, \emph{{Dimension-Six
  Terms in the Standard Model Lagrangian}},
  \href{https://doi.org/10.1007/JHEP10(2010)085}{\emph{JHEP} {\bfseries 10}
  (2010) 085}, [\href{https://arxiv.org/abs/1008.4884}{{\ttfamily 1008.4884}}].

\bibitem{Elias-Miro:2013eta}
J.~Elias-Miró, C.~Grojean, R.~S. Gupta and D.~Marzocca, \emph{{Scaling and
  tuning of EW and Higgs observables}},
  \href{https://doi.org/10.1007/JHEP05(2014)019}{\emph{JHEP} {\bfseries 05}
  (2014) 019}, [\href{https://arxiv.org/abs/1312.2928}{{\ttfamily 1312.2928}}].

\bibitem{Giudice:2016yja}
G.~F. Giudice and M.~McCullough, \emph{{A Clockwork Theory}},
  \href{https://doi.org/10.1007/JHEP02(2017)036}{\emph{JHEP} {\bfseries 02}
  (2017) 036}, [\href{https://arxiv.org/abs/1610.07962}{{\ttfamily
  1610.07962}}].

\bibitem{Jenkins:2013zja}
E.~E. Jenkins, A.~V. Manohar and M.~Trott, \emph{{Renormalization Group
  Evolution of the Standard Model Dimension Six Operators I: Formalism and
  lambda Dependence}},
  \href{https://doi.org/10.1007/JHEP10(2013)087}{\emph{JHEP} {\bfseries 10}
  (2013) 087}, [\href{https://arxiv.org/abs/1308.2627}{{\ttfamily 1308.2627}}].

\bibitem{Jenkins:2013wua}
E.~E. Jenkins, A.~V. Manohar and M.~Trott, \emph{{Renormalization Group
  Evolution of the Standard Model Dimension Six Operators II: Yukawa
  Dependence}}, \href{https://doi.org/10.1007/JHEP01(2014)035}{\emph{JHEP}
  {\bfseries 01} (2014) 035},
  [\href{https://arxiv.org/abs/1310.4838}{{\ttfamily 1310.4838}}].

\bibitem{Alonso:2013hga}
R.~Alonso, E.~E. Jenkins, A.~V. Manohar and M.~Trott, \emph{{Renormalization
  Group Evolution of the Standard Model Dimension Six Operators III: Gauge
  Coupling Dependence and Phenomenology}},
  \href{https://doi.org/10.1007/JHEP04(2014)159}{\emph{JHEP} {\bfseries 04}
  (2014) 159}, [\href{https://arxiv.org/abs/1312.2014}{{\ttfamily 1312.2014}}].

\bibitem{Wells:2015cre}
J.~D. Wells and Z.~Zhang, \emph{{Renormalization group evolution of the
  universal theories EFT}},
  \href{https://doi.org/10.1007/JHEP06(2016)122}{\emph{JHEP} {\bfseries 06}
  (2016) 122}, [\href{https://arxiv.org/abs/1512.03056}{{\ttfamily
  1512.03056}}].

\bibitem{Brivio:2014pfa}
I.~Brivio, O.~J.~P. Éboli, M.~B. Gavela, M.~C. Gonzalez-Garcia, L.~Merlo and
  S.~Rigolin, \emph{{Higgs ultraviolet softening}},
  \href{https://doi.org/10.1007/JHEP12(2014)004}{\emph{JHEP} {\bfseries 12}
  (2014) 004}, [\href{https://arxiv.org/abs/1405.5412}{{\ttfamily 1405.5412}}].

\bibitem{Buschmann:2014sia}
M.~Buschmann, D.~Goncalves, S.~Kuttimalai, M.~Schonherr, F.~Krauss and
  T.~Plehn, \emph{{Mass Effects in the Higgs-Gluon Coupling: Boosted vs
  Off-Shell Production}},
  \href{https://doi.org/10.1007/JHEP02(2015)038}{\emph{JHEP} {\bfseries 02}
  (2015) 038}, [\href{https://arxiv.org/abs/1410.5806}{{\ttfamily 1410.5806}}].

\bibitem{Corbett:2015ksa}
T.~Corbett, O.~J.~P. Eboli, D.~Goncalves, J.~Gonzalez-Fraile, T.~Plehn and
  M.~Rauch, \emph{{The Higgs Legacy of the LHC Run I}},
  \href{https://doi.org/10.1007/JHEP08(2015)156}{\emph{JHEP} {\bfseries 08}
  (2015) 156}, [\href{https://arxiv.org/abs/1505.05516}{{\ttfamily
  1505.05516}}].

\bibitem{Englert:2017aqb}
C.~Englert, R.~Kogler, H.~Schulz and M.~Spannowsky, \emph{{Higgs
  characterisation in the presence of theoretical uncertainties and invisible
  decays}}, \href{https://doi.org/10.1140/epjc/s10052-017-5366-8}{\emph{Eur.
  Phys. J.} {\bfseries C77} (2017) 789},
  [\href{https://arxiv.org/abs/1708.06355}{{\ttfamily 1708.06355}}].

\bibitem{Gori:2013mia}
S.~Gori and I.~Low, \emph{{Precision Higgs Measurements: Constraints from New
  Oblique Corrections}},
  \href{https://doi.org/10.1007/JHEP09(2013)151}{\emph{JHEP} {\bfseries 09}
  (2013) 151}, [\href{https://arxiv.org/abs/1307.0496}{{\ttfamily 1307.0496}}].

\bibitem{Henning:2014gca}
B.~Henning, X.~Lu and H.~Murayama, \emph{{What do precision Higgs measurements
  buy us?}},  \href{https://arxiv.org/abs/1404.1058}{{\ttfamily 1404.1058}}.

\bibitem{Henning:2014wua}
B.~Henning, X.~Lu and H.~Murayama, \emph{{How to use the Standard Model
  effective field theory}},
  \href{https://doi.org/10.1007/JHEP01(2016)023}{\emph{JHEP} {\bfseries 01}
  (2016) 023}, [\href{https://arxiv.org/abs/1412.1837}{{\ttfamily 1412.1837}}].

\bibitem{Huo:2015exa}
R.~Huo, \emph{{Standard Model Effective Field Theory: Integrating out
  Vector-Like Fermions}},
  \href{https://doi.org/10.1007/JHEP09(2015)037}{\emph{JHEP} {\bfseries 09}
  (2015) 037}, [\href{https://arxiv.org/abs/1506.00840}{{\ttfamily
  1506.00840}}].

\bibitem{ATLAS:2018doi}
{\scshape ATLAS} collaboration, T.~A. collaboration, \emph{{Combined
  measurements of Higgs boson production and decay using up to 80 fb$^{-1}$ of
  proton--proton collision data at $\sqrt{s}=$ 13 TeV collected with the ATLAS
  experiment}}, .

\bibitem{ATL-PHYS-PUB-2018-054}
{\scshape ATLAS Collaboration} collaboration, \emph{{Projections for
  measurements of Higgs boson cross sections, branching ratios, coupling
  parameters and mass with the ATLAS detector at the HL-LHC}},  Tech. Rep.
  ATL-PHYS-PUB-2018-054, CERN, Geneva, Dec, 2018.

\bibitem{Bevilacqua:2012em}
G.~Bevilacqua and M.~Worek, \emph{{Constraining BSM Physics at the LHC: Four
  top final states with NLO accuracy in perturbative QCD}},
  \href{https://doi.org/10.1007/JHEP07(2012)111}{\emph{JHEP} {\bfseries 07}
  (2012) 111}, [\href{https://arxiv.org/abs/1206.3064}{{\ttfamily 1206.3064}}].

\bibitem{Frederix:2017wme}
R.~Frederix, D.~Pagani and M.~Zaro, \emph{{Large NLO corrections in
  $t\bar{t}W^{\pm}$ and $t\bar{t}t\bar{t}$ hadroproduction from supposedly
  subleading EW contributions}},
  \href{https://doi.org/10.1007/JHEP02(2018)031}{\emph{JHEP} {\bfseries 02}
  (2018) 031}, [\href{https://arxiv.org/abs/1711.02116}{{\ttfamily
  1711.02116}}].

\bibitem{Alvarez:2016nrz}
E.~Alvarez, D.~A. Faroughy, J.~F. Kamenik, R.~Morales and A.~Szynkman,
  \emph{{Four Tops for LHC}},
  \href{https://doi.org/10.1016/j.nuclphysb.2016.11.024}{\emph{Nucl. Phys.}
  {\bfseries B915} (2017) 19--43},
  [\href{https://arxiv.org/abs/1611.05032}{{\ttfamily 1611.05032}}].

\bibitem{Aaboud:2018jsj}
{\scshape ATLAS} collaboration, M.~Aaboud et~al., \emph{{Search for
  four-top-quark production in the single-lepton and opposite-sign dilepton
  final states in pp collisions at $\sqrt{s}$ = 13 TeV with the ATLAS
  detector}}, {\emph{Submitted to: Phys. Rev.} (2018) },
  [\href{https://arxiv.org/abs/1811.02305}{{\ttfamily 1811.02305}}].

\bibitem{Sirunyan:2017roi}
{\scshape CMS} collaboration, A.~M. Sirunyan et~al., \emph{{Search for standard
  model production of four top quarks with same-sign and multilepton final
  states in proton–proton collisions at $\sqrt{s} = 13\,\text {TeV} $}},
  \href{https://doi.org/10.1140/epjc/s10052-018-5607-5}{\emph{Eur. Phys. J.}
  {\bfseries C78} (2018) 140},
  [\href{https://arxiv.org/abs/1710.10614}{{\ttfamily 1710.10614}}].

\bibitem{ATL-PHYS-PUB-2018-047}
{\scshape ATLAS Collaboration} collaboration, \emph{{HL-LHC prospects for the
  measurement of the Standard Model four-top-quark production cross-section}},
  Tech. Rep. ATL-PHYS-PUB-2018-047, CERN, Geneva, Dec, 2018.

\bibitem{CMS-PAS-FTR-18-031}
{\scshape CMS Collaboration} collaboration, \emph{{Projections of sensitivities
  for tttt production at HL-LHC and HE-LHC}},  Tech. Rep. CMS-PAS-FTR-18-031,
  CERN, Geneva, 1900.

\bibitem{Azzi:2019yne}
P.~Azzi et~al., \emph{{Standard Model Physics at the HL-LHC and HE-LHC}},
  \href{https://arxiv.org/abs/1902.04070}{{\ttfamily 1902.04070}}.

\bibitem{Alwall:2014hca}
J.~Alwall, R.~Frederix, S.~Frixione, V.~Hirschi, F.~Maltoni, O.~Mattelaer
  et~al., \emph{{The automated computation of tree-level and next-to-leading
  order differential cross sections, and their matching to parton shower
  simulations}}, \href{https://doi.org/10.1007/JHEP07(2014)079}{\emph{JHEP}
  {\bfseries 07} (2014) 079},
  [\href{https://arxiv.org/abs/1405.0301}{{\ttfamily 1405.0301}}].

\bibitem{Alloul:2013bka}
A.~Alloul, N.~D. Christensen, C.~Degrande, C.~Duhr and B.~Fuks,
  \emph{{FeynRules 2.0 - A complete toolbox for tree-level phenomenology}},
  \href{https://doi.org/10.1016/j.cpc.2014.04.012}{\emph{Comput. Phys. Commun.}
  {\bfseries 185} (2014) 2250--2300},
  [\href{https://arxiv.org/abs/1310.1921}{{\ttfamily 1310.1921}}].

\bibitem{Mangano:2018mur}
{\scshape FCC} collaboration, A.~Abada et~al., \emph{{Future Circular
  Collider}}, .

\bibitem{Polchinski:1998rr}
J.~Polchinski, \emph{{String theory. Vol. 2: Superstring theory and beyond}}.
\newblock Cambridge Monographs on Mathematical Physics. Cambridge University
  Press, 2007,
  \href{https://doi.org/10.1017/CBO9780511618123}{10.1017/CBO9780511618123}.

\bibitem{Qi}
F.~Qi, \emph{{A double inequality for the ratio of two non-zero neighbouring
  Bernoulli numbers}}, vol.~351.
\newblock Journal of Computational and Applied Mathematics, 2019.

\end{thebibliography}\endgroup

\end{document}